\documentclass[aps,prd,nofootinbib,groupedaddress,twocolumn,preprintnumbers,balancelastpage,longbibliography]{revtex4-1}

\usepackage{appendix}
\usepackage[utf8]{inputenc}
\usepackage{amsmath,amssymb,mathtools,bm}
\usepackage{graphicx, color, hepunits}
\usepackage[svgnames,x11names,dvipsnames]{xcolor}
\usepackage{float}
\usepackage{multirow}
 \usepackage{hyperref} %Automatically links \label and \ref commands; Always load last
\hypersetup{
    colorlinks=true,       % false: boxed links; true: colored links
    linkcolor=blue,        % color of internal links
    citecolor=blue,        % color of links to bibliography
    filecolor=magenta,     % color of file links
    urlcolor=blue          % color of external links
}
\usepackage[utf8]{inputenc}
\usepackage[english]{babel}
\usepackage{tabularx}

\definecolor{codegreen}{rgb}{0,0.6,0}
\definecolor{codegray}{rgb}{0.5,0.5,0.5}
\definecolor{codepurple}{rgb}{0.58,0,0.82}
\definecolor{backcolour}{rgb}{0.95,0.95,0.92}

\usepackage{listings}
\lstdefinestyle{mystyle}{
    backgroundcolor=\color{GhostWhite},
    commentstyle=\color{RoyalBlue},
    keywordstyle=\color{RawSienna},
    numberstyle=\tiny\color{black},
    stringstyle=\color{ForestGreen},
    basicstyle=\ttfamily\footnotesize,
    breakatwhitespace=false,         
    breaklines=true,                 
    captionpos=b,                    
    keepspaces=true,                 
    numbers=left,                    
    numbersep=5pt,                  
    showspaces=false,                
    showstringspaces=false,
    showtabs=false,                  
    tabsize=2
}
\begin{document}

\newcommand{\sledgehamr}{\texttt{sledgehamr}}
\newcommand{\sledgehamrP}{\texttt{sledgehamr}}
\newcommand{\amrex}{\texttt{AMReX}}
\newcommand{\MB}[1]{ \textcolor{blue}{(#1)} }

\newcommand\blfootnote[1]{%
  \begingroup
  \renewcommand\thefootnote{}\footnote{#1}%
  \addtocounter{footnote}{-1}%
  \endgroup
}

\title{Sledgehamr: Simulating scalar fields with adaptive mesh refinement}

\author{Malte Buschmann}
\email{m.s.a.buschmann@uva.nl}
\affiliation{GRAPPA Institute, Institute for Theoretical Physics Amsterdam, University of Amsterdam, Science Park 904, 1098 XH Amsterdam, The Netherlands}

\begin{abstract}
Understanding the nonlinear dynamics of coupled scalar fields often necessitates simulations on a 3D mesh. These simulations can be computationally expensive if a large scale separation is involved. A common solution is adaptive mesh refinement, which, however, greatly increases a simulation's complexity. 
In this work, we present \sledgehamr, an \amrex-based code package to make the simulation of coupled scalar fields on an adaptive mesh more accessible. Compatible with both GPU and CPU clusters, \sledgehamr\ offers a flexible and customizable framework. While the code had been primarily developed to evolve axion string networks, this framework enables various other applications, such as the study of gravitational waves sourced by the dynamics of scalar fields.
\end{abstract}
\maketitle

%\tableofcontents
\blfootnote{\sledgehamr is available on GitHub:\\\url{https://github.com/MSABuschmann/sledgehamr}}

\section{Introduction}
\label{sec:Introduction}
Dynamical systems frequently exhibit complex behavior across different spatial and temporal scales, but often, we are able to separate the large-scale dynamics from that at small scales.
But this is not always the case: when the microphysics nontrivially affects the macrophysics, one cannot disentangle both. This is particularly problematic for non-linear systems where one relies on numerical simulations to study their dynamics. Suddenly, we need large simulation volumes to capture the macrophysics but at the same time require high resolution to describe the microphysics correctly. The number of operations needed to evolve those systems can easily exceed our computational capabilities. 

\begin{figure*}[!htb]
\centering
\includegraphics[width=\textwidth]{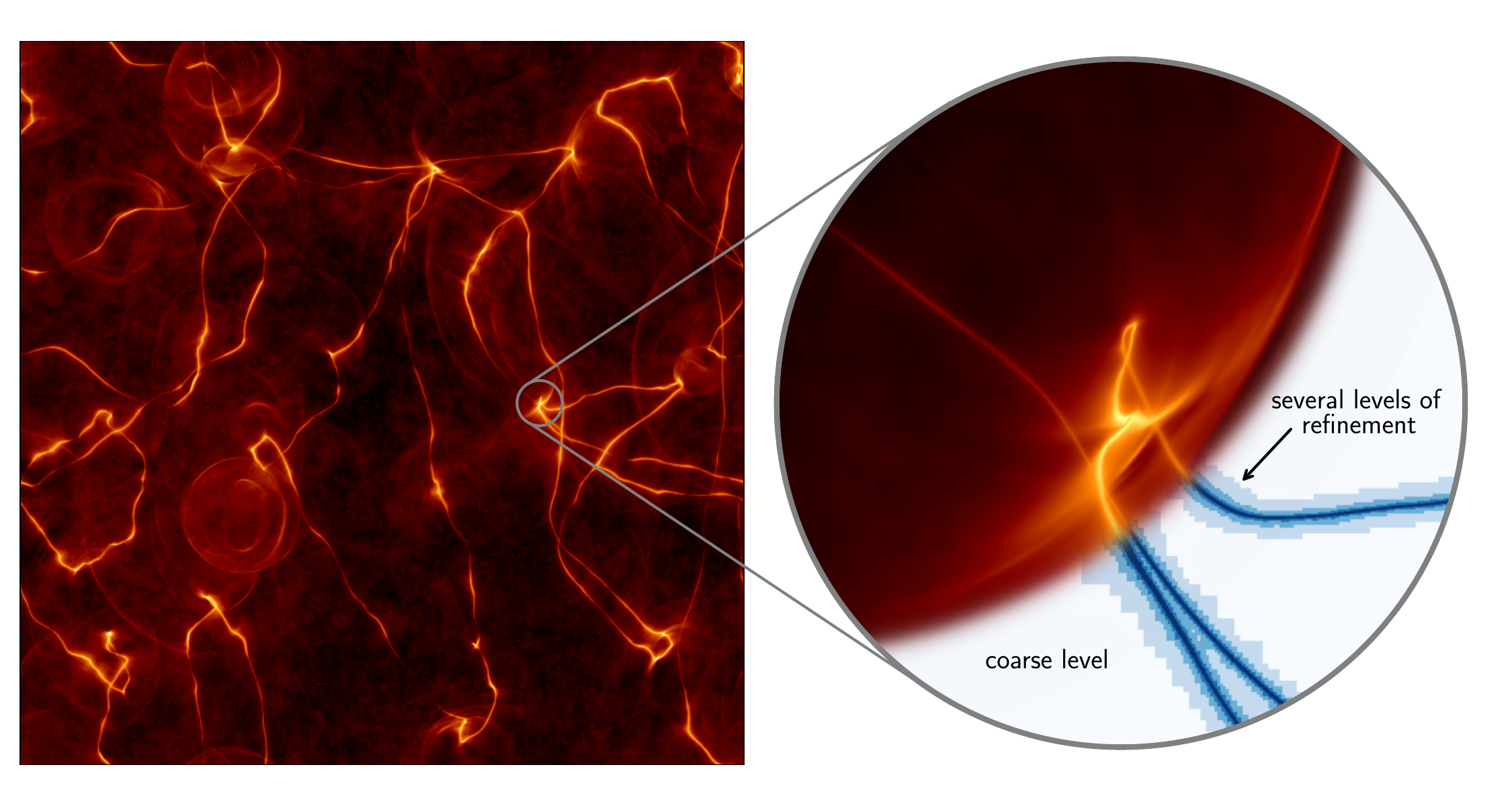}
\caption{
Illustration of axion strings based on a simulation presented in~\cite{AxionSim} using an older version of \sledgehamr. The left panel is a 3D$\rightarrow$2D projection (sum along the line of sight) of the energy density with a zoom-in in the right panel. Furthermore, the zoom-in shows the underlying refinement level coverage (finest level along the line of sight). In this particular snapshot, the axion string has a diameter of more than 5 orders of magnitude larger than its width. This makes them an ideal target for AMR simulations since refinement levels concentrated around the string can ensure that the dynamics within the core of the string are properly captured during the evolution without overresolving the voids between strings.
}
\label{fig:axion_string}
\end{figure*}

Plenty examples of such systems exist, particularly in particle cosmology. The thin walls of expanding vacuum bubbles during a first-order phase transition need to be resolved to correctly determine the spectrum of gravitational waves produced by bubble collision (see \textsl{e.g.}~\cite{Cutting:2018tjt, Cutting:2020nla, FOPT}). This is extremely challenging in (3+1) dimensions. Axion strings are another example. Axion strings are topological defects of a complex scalar field that appear after spontaneously breaking its underlying $U(1)$ symmetry. They form closed loops with a diameter that is roughly a Hubble length. The width of those axion strings, however, is many orders of magnitude smaller; see Fig.~\ref{fig:axion_string} for an illustration. This separation of scales is problematic: We need a simulation volume that is at least a Hubble volume large to fit a single axion string but at the same time need to resolve the much smaller string width to be able to evolve the system in time. Even though this scenario is well motivated and of interest, this has caused progress in understanding the dynamics to be slow and tedious (see for example~\cite{Gorghetto:2018myk, Gorghetto:2020qws, Gorghetto:2021fsn, Buschmann:2021sdq, Benabou:2023ghl, Saikawa:2024bta, Kim:2024wku} for a few recent simulations of axion strings).

This paper concerns physics scenarios that can be simulated on a 3D mesh, in particular systems of coupled scalar fields. Here, there is a well-known approach to alleviate the problem of a large scale separation, \textsl{adaptive mesh refinement} (AMR). Rather than having a large mesh with constant grid spacings, with the AMR technique selected parts of the simulation volume will have a higher spatial and temporal resolution than others. This approach works well if the effects of the microphysics are spatially localized. This is the case of axion strings, where high resolution is primarily needed only within the vicinity of the strings. 

Block-structured AMR was introduced in~\cite{Berger:1984zza} and proven to be extremely useful in the simulation of gas dynamics in both 2D~\cite{BERGER198964} and 3D~\cite{doi:10.1137/0915008}. Since then, it has been applied to various research fields and sparked the development of specialized AMR codes such as ENZO~\cite{Bryan_2014} and RAMSES~\cite{Teyssier:2001cp} for cosmological structure formation. Further examples for AMR implementations may include FLASH~\cite{Fryxell_2000} for thermonuclear flashes in and around compact stars, MAESTROeX~\cite{Fan_2019} for explosive astrophysical phenomena, and GRChombo~\cite{Clough:2015sqa} for general relativity simulations.

AMR adds vast amounts of complexity to the simulation, which can be a deterrent. Therefore, we present the code package \sledgehamr\ (\textbf{S}ca\textbf{L}ar fi\textbf{E}ld \textbf{D}ynamics \textbf{G}etting solv\textbf{E}d wit\textbf{H} \textbf{A}daptive \textbf{M}esh \textbf{R}efinement), with which AMR simulations of coupled scalar fields can be implemented with minimal coding. \sledgehamr\ aims to simplify the usage of AMR while still maintaining a large amount of flexibility to adapt to the specific needs of a particular project. 

\sledgehamr\ uses the \amrex\, software framework~\cite{AMReX_JOSS} internally, a generic AMR library for massively parallel, block-structured applications. As such, \sledgehamr\ is scalable with its MPI + tiled OpenMP approach and can run smoothly on both CPU and GPU clusters. Older unpublished versions of \sledgehamr\ have already been used in a few publications~\cite{Buschmann:2021sdq, Benabou:2023ghl, AxionSim, FOPT} where the code ran on up to 139,264 CPU cores and 1,024 GPUs.

\sledgehamr\ solves the (3+1)-dimensional equations of motion (EOMs) of coupled scalar fields with periodic boundary conditions. While the code was originally developed with early universe particle cosmology in mind, it has since undergone an extensive rewrite to accommodate any scenario involving coupled scalar fields. It further provides an interface to coevolve the tensor perturbations needed to calculate gravitational-wave production. 
Finding convergence criteria to determine which part of the simulation volume should be refined is as much art as it is science. \sledgehamr\ employs a data-driven technique using local truncation error estimates to obtain a multilevel AMR grid layout, although other physics-inspired criteria can be implemented as well.

We summarize the AMR terminology and the workflow of \sledgehamr\ in section~\ref{sec:AMR}. A minimal example with a detailed explanation is provided in section~\ref{sec:MinimalExample}. How a new physics scenario can be implemented in \sledgehamr\ from scratch is the topic of section~\ref{sec:implementation} with a summary of the gravitational-wave framework provided in section~\ref{sec:GW}. Section~\ref{sec:input} and~\ref{sec:output} describe the simulation input and output, respectively. We conclude in section~\ref{sec:discussion}.

\section{AMR}
\label{sec:AMR}
In this section, we want to go over some of the AMR aspects of the code and the general workflow. \sledgehamr\ is based on \amrex, a software framework for block-structured AMR. As such, the underlying grid is divided into rectangular boxes. To reduce computational overhead, the size of each box is always a multiple of the run-time parameter \texttt{amr.blocking\_factor}. The blocking factor needs to be a power of 2 and is typically set to 8, 16, or 32 cells. AMR implies we have multiple levels, where each extra refinement level decreases the grid spacing and time step size by a factor of 2. Only the coarse level covers the entire simulation volume whereas each finer level covers parts. How the individual levels are coordinated and advanced in time is described in detail in section~\ref{sec:integration}. 

The coarse-level grid spacing $\Delta x_{\ell=0}=L/N$ is set by specifying the simulation box length $L$ through \texttt{sim.L} and the coarse-level grid size $N$ through \texttt{amr.coarse\_level\_grid\_size}. The grid spacing $\Delta x_\ell$ at level $\ell$ is then given by
\begin{equation}
    \Delta x_\ell = \Delta x_{\ell=0}/2^\ell.
\end{equation}
The time step size $\Delta t_\ell$ is fixed by the Courant-Friedrichs-Lewy condition~\cite{Courant:1928lij} which dictates 
\begin{equation}
    \Delta t_\ell / \Delta x_\ell \leq C.
\end{equation}
$C$ is typically of order one and the ratio $\Delta t_\ell / \Delta x_\ell$ can be set through \texttt{sim.cfl}. This subsequently implies the time step size is decreased on each refinement level as well
\begin{equation}
    \Delta t_\ell = \Delta t_{\ell=0} / 2^\ell,
\label{eq:dtscaling}
\end{equation}
making this technically an adaptive time stepping scheme. Note, however, that this may pose certain challenges for some systems as the time step size is fixed for each level throughout the simulation. If part of a simulation converges poorly, the data-driven refinement criteria described in section~\ref{sec:tagging} will introduce extra refinement levels, reducing both $\Delta t$ and $\Delta x$ locally. This may not be optimal as a global and possibly temporary reduction of $\Delta t_\ell$ alone may be computationally favorable instead of introducing refinement levels locally. Some systems may thus benefit from a truly adaptive time stepping scheme where the global time step size is changed over time based on some time step stability criterion (see e.g.~\cite{Almgren_2010, Bryan_2014}). Such schemes can be implemented in \sledgehamr\ using the \texttt{BeforeTimestep} function described in section~\ref{sec:virtual} where the global time step size can be rescaled safely.

The refined region is periodically adjusted through a procedure known as \textsl{regridding}, hence \textsl{adaptive} mesh refinement.
Regridding is achieved by checking whether any coarse-level cell should be refined through the concept of cell tagging and then finding a mesh layout that covers all tagged cells. The details of cell tagging are described in section~\ref{sec:tagging}. The size of the regrid interval, $\Delta t_{\text{regrid},\ell=0}$, is set by \texttt{amr.regrid\_dt}. Note that because each refinement level decreases the time step size the regrid interval is also decreased for finer levels,
\begin{equation}
\Delta t_{\text{regrid},\ell} = \Delta t_{\text{regrid},\ell=0} / 2^\ell.
\end{equation}

\subsection{Integration scheme}
\label{sec:integration}
\begin{figure*}[!htb]
\centering
\includegraphics[width=\textwidth]{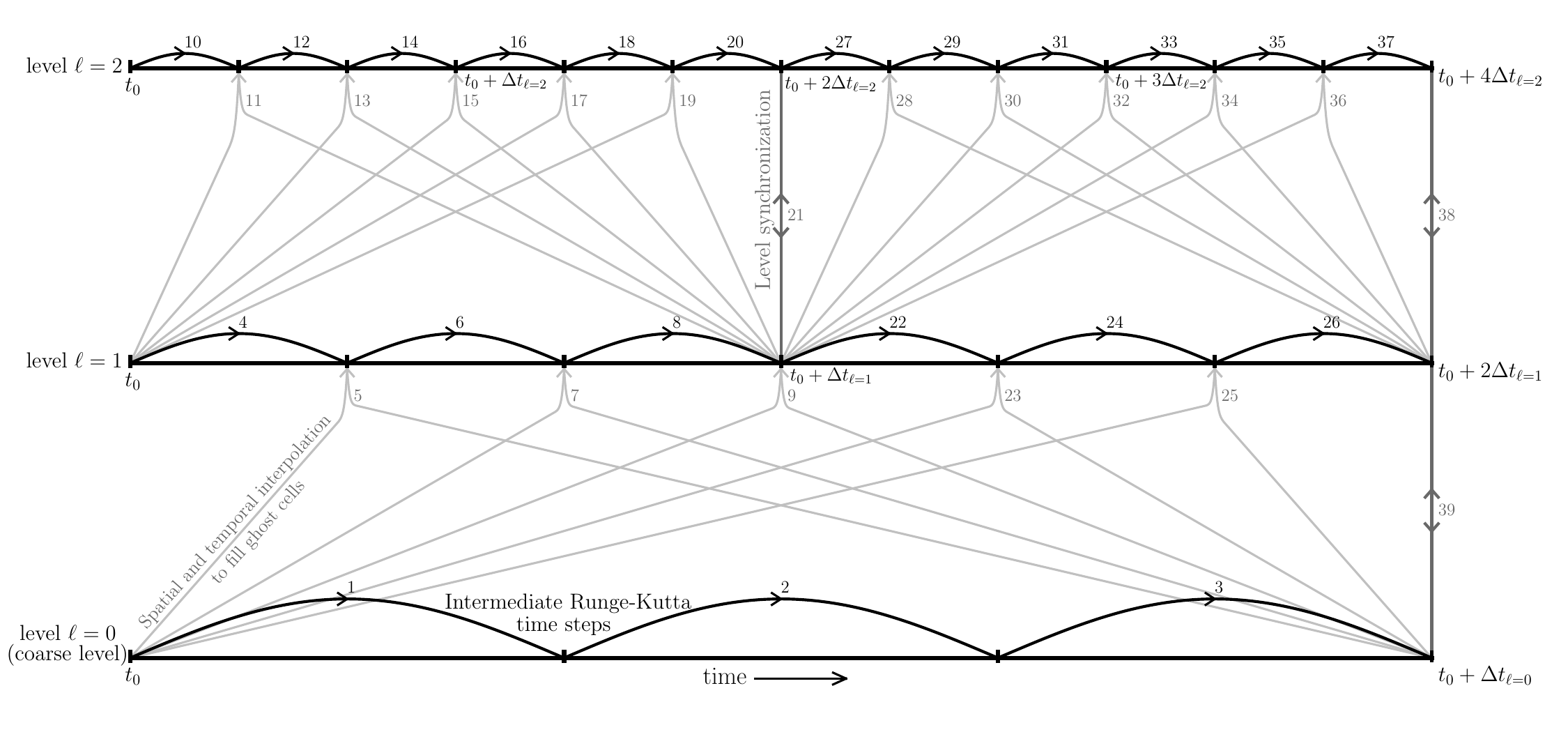}
\includegraphics[width=\textwidth]{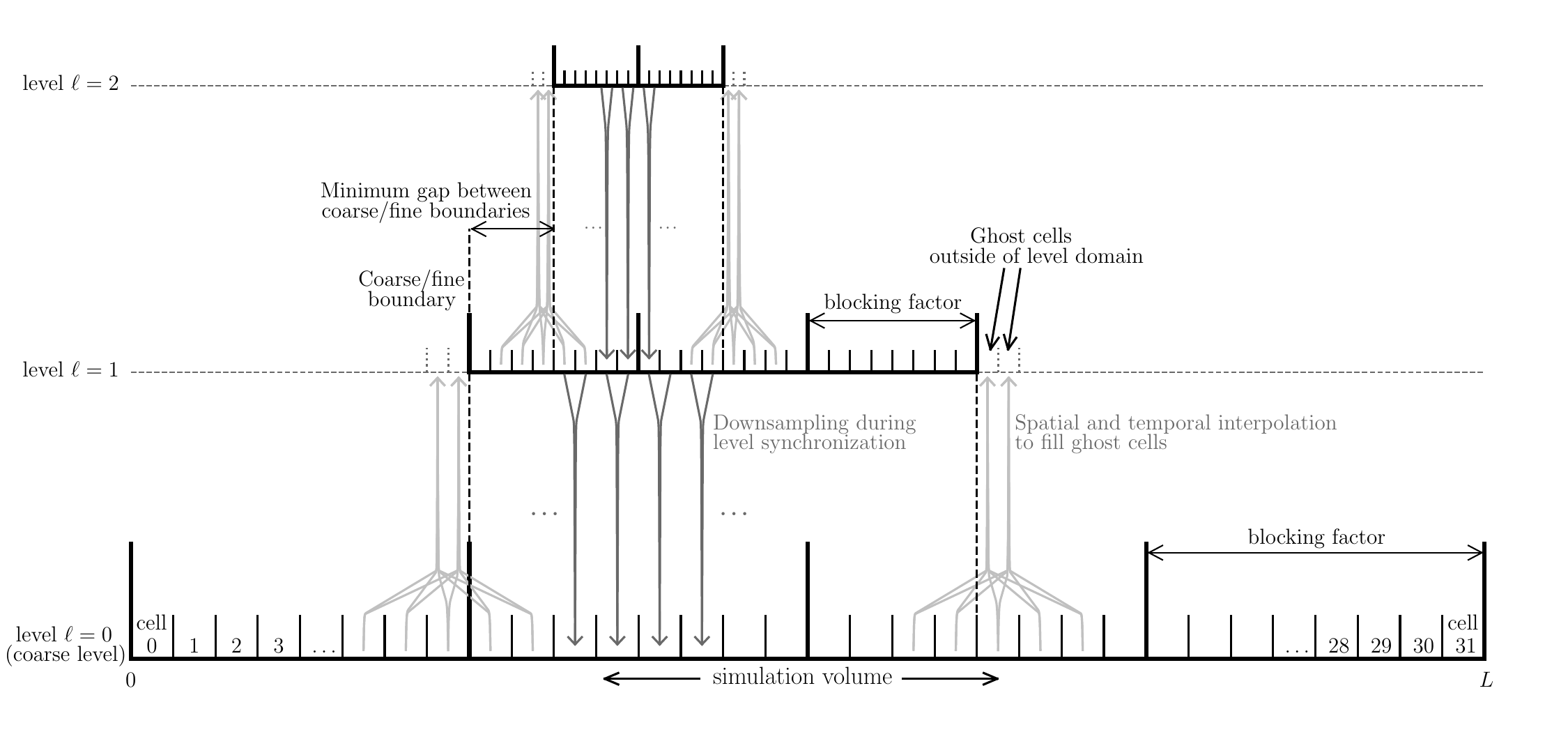}
\caption{
Illustration of the AMR workflow in the time domain \textsl{(top)} and the spatial domain \textsl{(bottom)} for a 1D simulation with two refinement levels and 32 coarse-level cells. The simulation advances in time using a recursive \textsl{time subcycling} algorithm, illustrated above using a three-stage integrator as an example. The numbers in the top panel indicate the order of operation during a single coarse-level time step. These operations consist of \textsl{(i)} the advancement of individual Runge-Kutta steps (1, 2, 3, \dots), \textsl{(ii)} the filling of fine-level ghost cells from a coarse level using spatial and temporal interpolation (5, 7, 9, \dots) and \textsl{(iii)} the synchronization of two adjacent levels (21, 38, 39).
Spatially, each grid is divided into boxes with a size that is a multiple of the \textsl{blocking factor}. A small number of ghost cells outside the valid domain of a fine level are filled through interpolation from the coarse level. To ensure proper coverage for this interpolation, a gap must exist between coarse/fine boundaries. Coarse-level cells that overlap with fine-level cells are overwritten through an averaging procedure during level synchronization.
}
\label{fig:workflow}
\end{figure*}

The integration scheme for an AMR simulation is slightly more complex than a standard fixed-lattice simulation as the various refinement levels need to be coordinated. We illustrate the general workflow for a single coarse-level time step in Fig.~\ref{fig:workflow}, both in the time and the spatial domain. We use a simple 1D simulation with 32 coarse-level cells and two refinement levels as an example though a 3D simulation works analogously. Furthermore, we use a three-stage integrator, such as a third-order Runge-Kutta scheme, for illustrative purposes, but this can be generalized to other integration schemes trivially. 

We use a time subcycling algorithm to advance the simulation in time, which evolves the individual levels recursively starting at the coarsest. First, the coarse level will be advanced fully regardless of any refinement levels from time $t_0$ to $t_0 + \Delta t_{\ell=0}$. This implies evolving also regions that are underresolved by the coarse level. This is intentional, as, otherwise, we would introduce nontrivial domain boundaries without a proper expression for the boundary conditions. These underresolved regions will be corrected at a later stage of the integration.

We can then move on to the next level, $\ell=1$, by advancing it by $\Delta t_{\ell=1} \equiv \Delta t_{\ell=0}/2$. Here, however, we need to be careful with boundary conditions, since, generally, the finer levels will not cover the entire simulation volume. But unlike the coarse level, the fine level has access to the underlying coarse-level data to fill in information outside its domain boundaries. To still be able to compute quantities like field gradients at the domain boundaries, we use the concept of \textsl{ghost cells}.

Ghost cells are a small number of cells, typically about one or two, that exist on the fine level just outside its domain. Ghost cells are not evolved like the rest of the fine level; instead, they are being interpolated from the underlying coarse level using a fourth-order spatial interpolation scheme. For this to work, we need to enforce a gap between coarse/fine boundaries, typically at least one fine-level blocking factor. Without such a gap, we may not be able to set ghost cells at level $\ell=2$ because the $\ell=1$ domain might not extend far enough for the interpolation.

Since we need to compute the boundary conditions at every intermediate integration step, ghost cells will need to be set regularly. However, since $\Delta t_{\ell=1} < \Delta t_{\ell=0}$, the coarse-level state does typically not exist for precisely the fine-level time at which we want to compute ghost cells. Therefore, we interpolate not only spatially but also temporally, using the coarse-level field state at $t_0$ and $t_0 + \Delta t_{\ell=0}$. Analogously, level $\ell=2$ will fill ghost cells from level $\ell=1$ using the states at $t_0$ and $t_0 + \Delta t_{\ell=1}$.

Once we recursively advance every level by their respective time step size $\Delta t_{\ell}$, we can start working backward by evolving the finest level a second time by $\Delta t_{\ell}$. The state at level $\ell$ and $\ell-1$ will now align in time, since $2\Delta t_{\ell} \equiv \Delta t_{\ell-1}$. This allows us to synchronize both levels, which entails downsampling the fine level to overwrite the underlying coarse-level cells. We also set fine-level ghost cells during this synchronization stage but no temporal interpolation is needed this time. We can also compute truncation error estimates (TEEs) during this synchronization; see section~\ref{sec:tagging}. 

This procedure can then be repeated until all levels are synchronized at $t_0 + \Delta t_{\ell=0}$. 

\subsection{Tagging cells for refinement}
\label{sec:tagging}
To determine which part of the simulation volume shall be refined, we \textsl{tag} cells. Every tagged cell will be refined, including a buffer region around it~\cite{120081}. The size of this buffer region is set by \texttt{amr.n\_error\_buf}. \sledgehamr\ can handle two different types of cell tagging criteria: data driven and problem specific.

The data-driven tagging criterion is readily available in \sledgehamr. The idea is that, if one advances identical states $\phi(x)$ at two different resolutions for a time $\Delta t$, the difference between the results, 
\begin{equation}
\Delta \phi_\text{cf}(x) = |\phi_\text{coarse}(x) - \phi_\text{fine}(x)|,
\end{equation}
informs us about their numerical convergence. If $\Delta \phi_\text{cf}$ is large, the convergence is bad, whereas if the difference is small, both states must have converged reasonably well. We will refer to $\Delta \phi_\text{cf}$ as the TEE. We can use $\Delta\phi_\text{cf}$ to identify poorly converging regions in our simulation and tag cells for refinement where $\Delta \phi_\text{cf}$ exceeds some threshold $\epsilon_\phi$. 

Of course, this requires us to simulate the same field at two resolutions, but luckily, this comes almost for free in an AMR simulation: A refined level $\ell$ is always nested within the coarser level $\ell-1$ and both levels are evolved independently\footnote{Except at the boundary region of level $\ell$ since ghost cells on level $\ell$ will be spatially and temporally interpolated from $\ell-1$ at each intermediate integrator step. This correlates level $\ell$ and $\ell-1$ which will be noticeable when computing $\Delta \phi_\text{cf}$ at these locations.} for a time $2\Delta t_{\ell} = \Delta t_{\ell-1}$ before they are synchronized again. Therefore, we have natural access to TEEs just before synchronization. 

This is known as the \textsl{self-shadow} method. Obviously, this method breaks down at the coarse level $\ell=0$ since we would naïvely not coevolve an even coarser level. To remedy this issue, we do indeed introduce such a coarsened version of the coarse level. We dub this the \textsl{shadow level} but only introduce it when needed shortly before cell tagging. Once the regrid has been performed, the shadow level is removed since we can trivially reconstruct it from the coarse level before the next regrid. The extra computational cost is relatively small since the shadow level is only evolved for a single time step $\Delta t_s=2\Delta t_{\ell=0}$ once every regridding interval. 

\begin{figure*}[!htb]
\centering
\includegraphics[width=\textwidth]{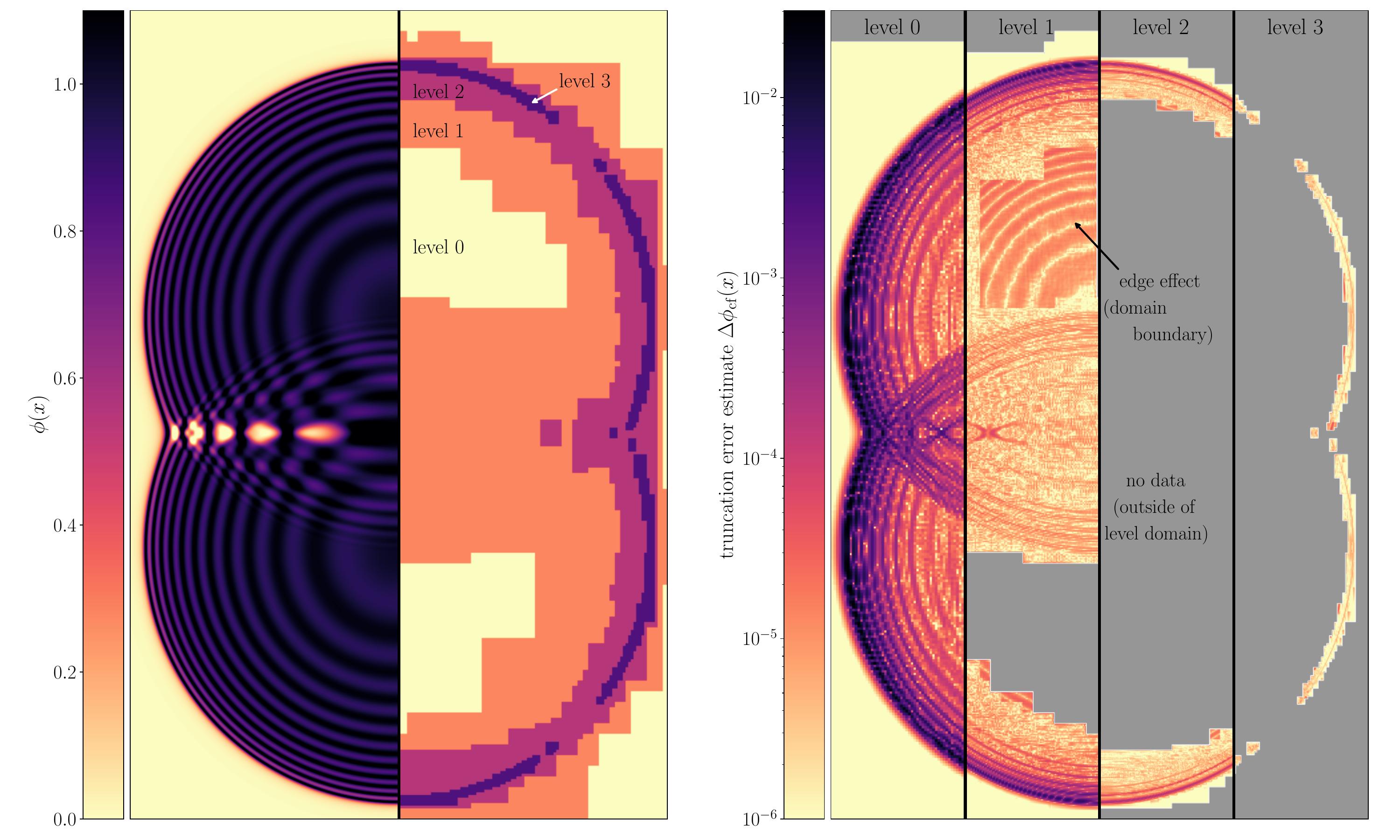}
\caption{
An example of truncation error estimates is illustrated using a slice through a 3D volume. For details of the simulation, see main text. \textsl{(Left)} Slice through one of the scalar field $\phi(x)$ on the left-hand side, and the respective refinement level coverage on the right-hand side. \textsl{(Right)} Corresponding truncation error estimates $\Delta\phi_\text{cf}(x)$ for all four levels. Gray areas contain no data as this part of the simulation volume is not covered by the respective level.
}
\label{fig:truncation_error}
\end{figure*}

We illustrate TEEs in Fig.~\ref{fig:truncation_error} using a snapshot of a (3+1)D simulation of a scalar field $\phi(x)$ undergoing a phase transition from the false vacuum at $\phi(x)=0$ to the true vacuum at $\phi(x)=1$. Specifically, Fig.~\ref{fig:truncation_error} shows a slice through two expanding and colliding bubbles of the true vacuum. The exact details of this scenario are not crucial for our illustration but can be found in~\cite{FOPT}. In this example, the outer bubble walls are sharp features that require high spatial resolution. This is captured by the fact that TEEs at the bubble walls are large at lower levels due to poor numerical convergence. Subsequently, most of the outer wall is refined down to level $\ell=3$. Note, however, how the TEE decreases for every extra refinement level until they are ultimately subthreshold; in this particular case, $\epsilon_\phi=10^{-3}$. We use this example to study the numerical convergence and computational resource consumption as a function of $\epsilon_\phi$ in the Appendix.

Furthermore, one might notice that there are certain boxlike patches separated by a sharp transition where the TEE is larger than in its immediate surroundings. In Fig.~\ref{fig:truncation_error} this is most obvious on refinement level 1. This is due to an edge effect as this area is right at the boundary of the level domain. In this particular case, level 1 does not extend farther into the third dimension in the positive direction. In these boundary cells, gradients appearing in the EOMs are not computed consistently using fine-level data but include ghost cells that are interpolated from the coarse level. The TEEs are therefore larger. This is not an issue, however, as these fine-level TEEs are by construction similar in scale to the TEEs of the coarse level. Since the coarse-level TEEs were subthreshold as otherwise, there would be no fine-level domain boundary at this location; the fine-level TEEs will also be subthreshold. 

The TEE threshold $\epsilon_\phi$ can be set through \texttt{amr.te\_crit}. Furthermore, the actual threshold is implemented as 
\begin{equation}
    f_\phi(\Delta \phi_\text{cf}) > \epsilon_\phi,
\end{equation} 
where $f_\phi(\Delta\phi_\text{cf})=\Delta\phi_\text{cf}$ by default. The function $f_\phi(\Delta\phi_\text{cf})$ can be changed, however, and we refer to section~\ref{sec:ModTE} for implementation details. Besides this data-driven method, it is also possible to implement extra problem-specific criteria to tag cells. How this is done in practice is described in section~\ref{sec:ProblemTagging}.

\subsection{Local versus global regrid}
\label{sec:regrid}
One unique aspect of \sledgehamr\ we would like to highlight is the difference between \textsl{global} and \textsl{local} regrids. The \amrex\ framework provides a regridding method where cells are tagged for refinement, and then, a Berger-Rigoutsos clustering algorithm is applied to determine the optimal grid layout that encompasses all tagged cells; see \cite{120081}. We refer to this technique as \textsl{global} regrid as it clusters all tagged cells globally. A global regrid is sufficient in many use cases as it provides an optimized load-balanced grid within a reasonable amount of computing time. 

There are, however, cases where a global regrid can become a computational bottleneck. For very large simulation volumes, such as those in~\cite{AxionSim}, the grid has to be chopped into an enormous amount of smaller boxes. Determining the optimal layout in this case can take a significant amount of computing time. \sledgehamr\ provides a faster alternative to the global regrid, which we will refer to as a \textsl{local} regrid. 

During a local regrid, cells will be tagged for refinement as usual. However, rather than clustering them to determine a new grid layout, a local regrid will leave the existing grid as is. Instead, it identifies tagged cells that are too close to coarse/fine boundaries and simply expands the existing grid around them if needed. Here, it is assumed that any feature cannot travel farther than \texttt{n\_error\_buf} cells in between regrids, and a cell is therefore considered too close to a grid boundary if it is closer than \texttt{n\_error\_buf} cells. As this entails mostly local operations, the algorithm could be parallelized more effectively and thus speed up the regrid.

This algorithm is particularly efficient if the majority of features that require refinement travel slower than their theoretical maximum. For example, in the case of axion strings, any string segment could theoretically move as fast as the speed of light, and near relatively rare string reconnections, they will indeed saturate that limit~(\cite{Buschmann:2021sdq, AxionSim}). This forced the conservative choice of a rather large buffer region \texttt{n\_error\_buf} to ensure even the fastest string segment would not be able to leave the refined grid. 
However, the majority of string segments will move much slower and as such are surrounded by a far larger buffer region than it needs to be.  During a local regrid, it will take several regridding intervals for most string segments to get close enough to the grid boundary to trigger extra refinement. This means much of the grid will stay untouched during a local regrid for several regridding intervals. In~\cite{AxionSim}, a local regrid took a few seconds compared to a few minutes for the global regrid, alleviating the bottleneck. 

It is important to note that the local regrid comes with several drawbacks despite it being incredibly fast. Most importantly, it only ever increases the refined area and never removes any. Therefore, the local regrid can only ever be used in conjunction with an occasional global regrid to reoptimize the grid layout.

Fortunately, the local regrid module has a built-in logic to automatically determine when a global regrid should be performed instead of a local regrid. It does that by keeping track of how much extra volume has been added by all local regrids since the last global regrid. If this fraction $dV_\ell$ exceeds some threshold $\epsilon_{s}$ on any refinement level $\ell$, it requests a global regrid at the next opportunity starting at the lowest level where $dV_\ell \geq \epsilon_{w}$. Typically, we have $\epsilon_w < \epsilon_s$ to avoid lower levels triggering a global regrid when higher refinement levels were globally regridded just a moment prior. By default, we have $\epsilon_{s}=10\%$ and $\epsilon_{w}=5\%$, although these can be adjusted in the inputs file through \texttt{amr.volume\_threshold\_strong = $\epsilon_s$} and \texttt{amr.volume\_threshold\_weak = $\epsilon_w$}. Furthermore, the maximum number of consecutive local regrids without a global regrid can be capped by setting \texttt{amr.max\_local\_regrids}.

There is one more difference between a global and a local regrid that we would like to highlight. If a global regrid is performed at level $\ell$, the current \amrex\ implementation will only modify level $\ell$ and higher. Coarser levels will be left untouched. This can become an issue at times as level $\ell$ needs to be always nested within level $\ell-1$ with at least a \texttt{blocking\_size}'d wide gap between the boundary of level $\ell-1$ and the boundary of level $\ell$. If a cell too close to the boundary of level $\ell-1$ is tagged, the algorithm will not be able to refine this cell, or otherwise, it will violate nesting requirements. This caveat is not present in a local regrid. Here, if a local regrid is performed at level $\ell$, the algorithm will also expand lower levels solely to ensure nesting.

\subsection{Kreiss-Oliger dissipation}
\label{sec:KO}
AMR mixes grids of different resolutions, which can be prone to numerical instabilities introduced by the finite-difference method at the respective level boundaries. This can manifest, e.g., through spurious high-frequency modes though the severity is highly problem specific (see e.g.~\cite{LI20103139} for a discussion). In cases where these artifacts become a concern, a common mitigation strategy is the introduction of an extra artificial dissipation term. \sledgehamr\ implements Kreiss-Oliger dissipation~\cite{KOdissipation} by adding the small-scale operator 
\begin{equation}
    Q_\ell = (-1)^{\frac{p+3}{2}}\epsilon_\text{KO} \frac{\Delta x_\ell^p}{2^{p+1}}\left(\frac{\partial^{(p+1)}}{\partial x^{(p+1)}}+\frac{\partial^{(p+1)}}{\partial y^{(p+1)}}+\frac{\partial^{(p+1)}}{\partial z^{(p+1)}}\right)
\end{equation}
to the evolution of each scalar field component. Here, $\Delta x_\ell$ is the grid spacing at level $\ell$, $p$ is the dissipation order, $\partial^{(p+1)}/\partial x^{(p+1)}$ is the $(p+1)$-\textsl{st} spatial derivative, and $\epsilon_\text{KO}$ is the overall dissipation strength. Kreiss-Oliger dissipation can be switched on by setting the order $p$ through \texttt{sim.dissipation\_order} and the strength $\epsilon_\text{KO}$ through \texttt{sim.dissipation\_strength}. The appropriate size of $\epsilon_\text{KO}$ is somewhat problem specific and is sometimes not needed at all if the scenario at hand naturally dissipates energy. As we demonstrate in appendix \ref{sec:AppKO}, however, $\epsilon_\text{KO}\sim\mathcal{O}(10^{-2})$ is sufficient in many cases.

\section{Minimal example}
\label{sec:MinimalExample}
\subsection{Implementation}
In this chapter, we illustrate how \sledgehamr\ can be used to simulate a simple physics scenario with AMR. We use axion strings as an example, which require us to simulate a complex scalar field $\Phi$ with the Lagrangian
\begin{equation}
 \mathcal{L} = |\partial \Phi|^2 - \left(|\Phi|^2 - \frac{f_a}{2}\right)^2 - \frac{T^2}{3}|\Phi|^2.
\end{equation}

From this Lagrangian, we can derive the EOMs for the two (dimensionless) components of $\Phi = \frac{f_a}{\sqrt{2}}(\psi_1 - i\psi_2)$ which yields two coupled second-order differential equations
\begin{equation}
    \frac{\partial^2\psi_i}{d\eta^2} + \frac{2}{\eta}\frac{\partial\psi_i}{d\eta} - \nabla^2\psi_i + \eta^2\psi_i(\psi_1^2+\psi_2^2-1) + a\psi_i=0.
\end{equation}
Here, $\eta$ is (comoving) time, $\nabla^2=\partial^2/dx^2+\partial^2/dy^2+\partial^2/dz^2$ is the Laplace operator, and $a$ is some constant.

See~\cite{Buschmann:2021sdq} for more details and derivations. 
We will further decompose both second-order equations into four first-order equations. We then have a total of 4 scalar degrees of freedom, $\psi_1$ and $\psi_2$, and their conjugate momenta $\Pi_1 \equiv \partial\psi_1/d\eta$ and $\Pi_2 \equiv \partial\psi_2/d\eta$ with respective EOMs
\begin{equation}
    \frac{\partial\psi_i}{d\eta} = \Pi_i,
\label{eq:EOM1}
\end{equation}
and
\begin{equation}
    \frac{\partial\Pi_i}{d\eta} = -\frac{2}{\eta}\Pi_i + \nabla^2\psi_i - \eta^2\psi_i(\psi_1^2+\psi_2^2-1) - a\psi_i.
\label{eq:EOM2}
\end{equation}

To simulate this set of four equations using \sledgehamr\ we need to first start a new project by creating a header file \texttt{MinimalExample.h} inside the folder \texttt{sledgehamr/projects/MinimalExample/}. We chose \texttt{MinimalExample} for our project name, but this could have been anything as long as the header file name matches the folder it is contained in.  \texttt{MinimalExample.h} then consists of the following code:

\begin{lstlisting}[language=c++]
#pragma once
#include <sledgehamr.h>
namespace MinimalExample {

SLEDGEHAMR_ADD_SCALARS(Psi1, Psi2)
SLEDGEHAMR_ADD_CONJUGATE_MOMENTA(Pi1, Pi2)

AMREX_GPU_DEVICE AMREX_FORCE_INLINE
void Rhs(
    const amrex::Array4<double>& rhs,
    const amrex::Array4<const double>& state,
    const int i, const int j, const int k,
    const int lev, const double time, 
    const double dt, const double dx,
    const double* params) {
    
    // Fetch field values.
    double Psi1 = state(i, j, k, Scalar::Psi1);
    double Psi2 = state(i, j, k, Scalar::Psi2);
    double Pi1  = state(i, j, k, Scalar::Pi1);
    double Pi2  = state(i, j, k, Scalar::Pi2);
    double eta = time;

    // Compute Laplacians.
    constexpr int order = 2;
    double laplacian_Psi1 = 
        sledgehamr::utils::Laplacian<order>(
            state, i,j,k, Scalar::Psi1, dx*dx);
    double laplacian_Psi2 = 
        sledgehamr::utils::Laplacian<order>(
            state, i,j,k, Scalar::Psi2, dx*dx);

    // Compute potential.
    const int a = 0.56233;
    double pot = 
        eta*eta*(Psi1*Psi1 + Psi2*Psi2 -1) + a;

    // Now compute the EOMs.
    rhs(i, j, k, Scalar::Psi1) =  Pi1;
    rhs(i, j, k, Scalar::Psi2) =  Pi2;
    rhs(i, j, k, Scalar::Pi1)  = 
        -Pi1*2./eta + laplacian_Psi1 - Psi1*pot;
    rhs(i, j, k, Scalar::Pi2)  =
        -Pi2*2./eta + laplacian_Psi2 - Psi2*pot;
}

SLEDGEHAMR_FINISH_SETUP
class MinimalExample 
    : public sledgehamr::Sledgehamr {
 public:
  SLEDGEHAMR_INITIALIZE_PROJECT(MinimalExample)
};

}; // namespace
\end{lstlisting}
This code is all that is required to simulate axion strings on an adaptive mesh using \sledgehamr. The first three lines consist of the standard header guard, an \texttt{\#include} statement, and the beginning of a namespace. It is crucial to encapsulate our entire project code within a namespace that matches our project name to avoid unwanted conflicts with other projects. We then use the \texttt{SLEDGEHAMR\_ADD\_SCALARS} and \texttt{SLEDGEHAMR\_ADD\_CONJUGATE\_MOMENTA} macros to name the fields we want to simulate, in our case $\psi_{1,2}$ and $\Pi_{1,2}$. These macros will set the symbols \texttt{Scalar::Psi1}, etc., which we will use to access individual field components later on.

Next, we have the opportunity to define \textsl{kernel functions}. Kernel functions are functions that operate on a single grid cell and are thus called numerous times during a simulation. If GPU support is enabled during compilation, kernel functions will be executed on GPUs.
For consistency, we will refer to kernel functions as kernel function even if \sledgehamr\ has not been compiled with GPU support.
There is only one kernel function that has to be defined in each project, and it is the \texttt{Rhs} function. As its name suggests, its purpose is to compute the right-hand side of the EOMs. There exist other optional kernel functions that modify the behavior of \sledgehamr\ but those are discussed in section~\ref{sec:implementation}.

The \texttt{Rhs} function has various function arguments from which the EOMs can be computed. First, we have the two containers \texttt{rhs} and \texttt{state}. \texttt{state} contains the current field state whereas \texttt{rhs} is to be filled with the final result of the right-hand side of the EOMs in Eq.~\eqref{eq:EOM1} and~\eqref{eq:EOM2}. The indices $i$, $j$, and $k$ are the cell indices for which we want to compute the EOMs. \texttt{state} will only contain valid field values at this particular cell, as well as neighboring cells up to a distance defined by the run-time parameter \texttt{amr.n\_ghost\_cells}. This way \texttt{state} will contain just enough information to compute the local field gradients that appear in the EOMs.

Other function arguments include the current refinement level \texttt{lev}, current simulation time \texttt{time}, time step size \texttt{dt}, and grid spacing \texttt{dx}. In addition, there is the option to pass other problem-specific parameters to this function, as described in section~\ref{sec:virtual}. Those would be available through \texttt{params}, however, in this particular example, we do not need to use this functionality.

In the function body, we compute the EOMs by first fetching all field values at our cell from the \texttt{state} container.
We then determine the laplacians $\nabla^2\psi_{1,2}$ using one of the \sledgehamr\ utility functions (see appendix~\ref{sec:utility}). Here, \texttt{order} describes how many neighboring cells in each direction should be used during the calculation. Therefore, \texttt{order} must always be smaller or equal to the number of ghost cells. We then simply save the final result by setting the values of the \texttt{rhs} container for all four field components according to Eq.~\eqref{eq:EOM1} and~\eqref{eq:EOM2}.

We finish our project by introducing a derived class \texttt{MinimalExample}, where, again, its name must match our project name. The macro \texttt{SLEDGEHAMR\_FINISH\_SETUP} between kernel functions and class declaration together with the macro \texttt{SLEDGEHAMR\_INITIALIZE\_PROJECT(\$ProjectName)} within the class body ensures that our project is properly linked with \sledgehamr. The class \texttt{MinimalExample} is empty in our particular example and seems of little use. However, its base \texttt{sledgehamr::Sledgehamr} contains various virtual functions that can be overridden to customize the simulation workflow. More on this in section~\ref{sec:implementation}.

\subsection{Compilation and Run-Time Parameters}
We are now ready to compile our project using a makefile such as
\begin{lstlisting}[language=make]
AMREX_HOME      ?= /path/to/AMReX
SLEDGEHAMR_HOME ?= /path/to/sledgehamr

COMP      = gnu
USE_CUDA  = TRUE

include $(SLEDGEHAMR_HOME)/Make.sledgehamr
\end{lstlisting}
and executing \texttt{make -j 6}. In this makefile we provide file paths to the \amrex\ and \sledgehamr\ GitHub repositories, define a compiler environment, and whether we want to compile with GPU support. The created binary can then be run using, \textsl{e.g.}, \texttt{srun}. Often something like this will work, but your mileage may vary depending on your system architecture:
\begin{lstlisting}[language=sh]
export SLURM_CPU_BIND="cores"
export OMP_PLACES=threads
export OMP_PROC_BIND=spread
export OMP_NUM_THREADS=16
srun main3d.gnu.MPI.OMP.CUDA.ex inputs
\end{lstlisting}
In this particular case, \sledgehamr\ was compiled with MPI, OpenMP, and CUDA support. Note how the executable expects a file called \texttt{inputs}. This file will contain all run-time parameters such as start and end time, grid size, box length, and much more. Here is an example:

\begin{lstlisting}[language=xml]
# ----------------- Select project                                              
project.name = MinimalExample 
    
# ----------------- Simulation parameters                                       
sim.t_start = 0.1                                                               
sim.t_end   = 5.5                                                                
sim.L       = 15                                                                
sim.cfl     = 0.3
    
# ----------------- Integrator                                                  
integrator.type = 10
                                                      
# ----------------- AMR parameters
amr.coarse_level_grid_size  = 256
amr.blocking_factor         = 8
amr.nghost                  = 2
amr.max_refinement_levels   = 2 
amr.n_error_buf             = 3
amr.regrid_dt               = 0.2
                            
amr.te_crit = 1e-2
                        
# ----------------- Input parameters
input.initial_state = initial_state.hdf5

# ----------------- Output settings
output.output_folder       = output
output.slices.interval     = 0.5
output.coarse_box.interval = 3 
\end{lstlisting}

A full list of all input parameters including an explanation can be found in appendix~\ref{sec:parameters}. The \sledgehamr\ GitHub repository provides a \texttt{README} and a Jupyter notebook with detailed instructions for how to compile, run, and analyze the example.

\section{Customizing a new project}
\label{sec:implementation}
This chapter is about how one can start a new project and customize the simulation workflow to adapt to the problem-specific needs of the project.

The easiest way to start a new project from scratch is to create a copy of the \texttt{MinimalExample} folder under $\texttt{sledgehamr/projects/}$ and replace every instance of \texttt{MinimalExample} with the new project name. And that is it. \sledgehamr\ will automatically detect that a new project has been created and will include it the next time it is compiled.

Besides the obvious modifications such as changing the number of field components in the \texttt{SLEDGEHAMR\_ADD\_SCALARS} and \texttt{SLEDGEHAMR\_ADD\_CONJUGATE\_MOMENTA} macro calls, and adapting the EOMs in the \texttt{Rhs} kernel function, various other options exist to change the simulation workflow. These modifications are typically done either \textsl{(i)} through the implementation of specific kernel functions or \textsl{(ii)} through overriding virtual functions of the project class. Several of these techniques are demonstrated in the \texttt{NextToMinimalExample} provided with the GitHub repository.

\subsection{Virtual project functions}
\label{sec:virtual}
 The project class is a class derived from \texttt{sledgehamr::Sledgehamr}, as seen in the example in section~\ref{sec:MinimalExample}. As such, it does not only inherit access to the raw data through member functions such as \texttt{GetLevelData(const int lev)}\footnote{For a full list of available functions, see\\\url{https://msabuschmann.github.io/sledgehamr/}} but also can modify the workflow through the overriding of virtual functions.
 
The most commonly overridden virtual function is \texttt{void Init()}, which will be called at the end of the initialization phase. Here, we could perform extra modifications to the initial state, parse project-specific input parameters from the \texttt{inputs} file using \texttt{amrex::ParmParse}, or even add extra custom output types (see section~\ref{sec:CustomOutputs}). At this stage, it would also be possible to set up a custom routine to parse an initial state with multiple refinement levels by initializing the state data manually (see \cite{AMReX_JOSS} for reference).

Another useful virtual function is 
\begin{lstlisting}[language=c++]
void SetParamsRhs( 
    std::vector<double>& params,
    const double time, const int lev)
\end{lstlisting}
which gives us access to an empty \texttt{params} vector. We can fill this vector with data that will then become accessible in the \texttt{Rhs} kernel function through the \texttt{const double* params} argument. \texttt{SetParamsRhs} will be called before every time step, so \texttt{params} can be filled with time-dependent data.

A common use case would be constants in the EOMs. For example, we could add a new parameter $c$ to the inputs file, \texttt{project.c = 1}, and parse this parameter in the \texttt{Init()} function through
\begin{lstlisting}[language=c++]
amrex::ParmParse pp("project"); 
pp.get("c", c); 
\end{lstlisting}
By then adding \texttt{c} to the \texttt{params} vector using \texttt{params.push\_back(c)} in the \texttt{SetParamsRhs} function, we can fetch it in the \texttt{Rhs} kernel function through \texttt{double c = params[0]}.

Further virtual functions at our disposal are 
\begin{lstlisting}[language=c++]
bool CreateLevelIf(const int lev,
                   const double time);
bool StopRunning(const double time);
void BeforeTimestep(const double time); 
\end{lstlisting}
The \texttt{CreateLevelIf} function is an extra criterion for whether we want to introduce a refinement level $\ell=$\texttt{lev}. By default, any refinement level is allowed to be created at any time; however, with this virtual function, we could delay its creation based on some problem-specific measure. Similarly, \texttt{StopRunning} allows us to override the stopping criterion of the simulation, which by default is $t \geq t_\text{end}$. Lastly, \texttt{BeforeTimestep} is a function that will be executed before every coarse-level time step. Its function is versatile and could be, for example, used to update auxiliary data periodically. Another use case of \texttt{BeforeTimestep} is to implement a custom adaptive time stepping scheme by simply changing the \texttt{dt} member variable of the parent class. Since this function is only called at the beginning of a coarse step, it is safe to change $\texttt{dt}[\ell]\equiv\Delta t_\ell$ here without accidentally desyncing refinement levels, as long as Eq.~\ref{eq:dtscaling} is still satisfied. As time step stability criteria are problem specific it is up to the user to implement a scheme that fits their system.

\subsection{Problem-specific cell tagging}
\label{sec:ProblemTagging}
Occasionally, tagging individual cells for refinement through TEEs might not be sufficient, or we merely want to complement it with some other criterion. A problem-specific tagging criterion can be added to a project by simply defining the following kernel function:
\begin{lstlisting}[language=c++]
template<>
AMREX_GPU_HOST_DEVICE AMREX_FORCE_INLINE
bool TagCellForRefinement<true>(
    const amrex::Array4<const double>& state,           
    const int i, const int j, const int k, 
    const int lev, const double time, 
    const double dt, const double dx,
    const double* params) { 
    \\ return true or false ... 
}
\end{lstlisting}
Each function argument is analog to those of the \texttt{Rhs} kernel, and the \texttt{params} argument can be set by overriding the corresponding virtual \texttt{SetParamsTagCellForRefinement} function. Nothing else needs to be done\footnote{The above code snippet is technically a template specialization thus it will be automatically integrated.} as long as this kernel function definition is added to the correct space in between the \texttt{SLEDGEHAMR\_ADD\_CONJUGATE\_MOMENTA} and \texttt{SLEDGEHAMR\_FINISH\_SETUP} macro calls; see the example in section~\ref{sec:MinimalExample}. Cell tagging through TEEs can be used simultaneously, and a cell will be tagged if either of the two methods tag it.

\subsection{Modifying truncation error estimates}
\label{sec:ModTE}
Cells on level $\ell$ are tagged for refinement through TEEs if 
\begin{equation}
    f_\phi(\Delta\phi_{\text{cf},\ell}) > \epsilon_\phi,
\end{equation} where $\Delta\phi_{\text{cf},\ell}(x) = |\phi_\ell(x) - \phi_{\ell-1}(x)|$ is the difference of the scalar field $\phi(x)$ between the coarse level $\ell-1$ and fine level $\ell$ after evolving them independently for $2\Delta t_\ell\equiv \Delta t_{\ell-1}$. By default, we have $f_\phi(\Delta\phi_\text{cf})=\Delta\phi_\text{cf}$ and $\epsilon_\phi$ can be set in the \texttt{inputs} file through \texttt{amr.te\_crit\_phi} (or \texttt{amr.te\_crit} for all scalar field components simultaneously). The function $f_\phi(\Delta\phi_\text{cf})$ can be modified by adding the following kernel function:
\begin{lstlisting}[language=c++]
template<> 
AMREX_GPU_HOST_DEVICE AMREX_FORCE_INLINE                           
double TruncationModifier<Scalar::phi>(
    const amrex::Array4<const double>& state,          
    const int i, const int j, const int k,
    const int lev, const double time, 
    const double dt, const double dx,                    
    const double truncation_error,
    const double* params) {
    // return f_\phi(\Delta\phi_{cf})
}
\end{lstlisting}
Note that this is an explicit template specialization for \texttt{Scalar::phi}, so we can implement a different function for each scalar field component separately. All function arguments are analog to the \texttt{Rhs} kernel function, with the addition of \texttt{truncation\_error} $\equiv\Delta\phi_\text{cf}(x)$. 

Similarly, just like for the \texttt{Rhs} kernel function, we can set \texttt{params} through the function \texttt{SetParamsTruncationModifier}. This can be particularly useful as $\Delta\phi_\text{cf}$ is an absolute quantity and thus overall field amplitude might matter. One could use \texttt{SetParamsTruncationModifier} to compute, for example, a standard deviation and pass it through \texttt{params} to normalize $\Delta\phi_\text{cf}$ to the TEE threshold $\epsilon_\phi$.

\section{Gravitational Waves}
\label{sec:GW}
The motivation for studying the evolution of coupled scalar fields is often the gravitational waves sourced by their dynamics. In \sledgehamr, it is possible to simulate gravitational-wave production relatively easily; we simply need to toggle them on with \texttt{sim.gravitational\_waves=1} and provide the corresponding EOMs. Here, we use the auxiliary field method~\cite{Figueroa:2011ye} where the transverse-traceless gauge-invariant metric perturbation $h_{ij}$ are evolved through auxiliary fields $u_{ij}$. Both are related through
\begin{equation}
    h_{ij}(t, \mathbf{k}) = \Lambda_{ij,lm}(\mathbf{k})u_{ij}(t, \mathbf{k})
\end{equation}
where the object $\Lambda_{ij,lm}$ is constructed from the projector
\begin{equation}
P_{ij}(\mathbf{k})=\delta_{ij} - \hat{k}_i^\text{eff}\hat{k}_j^\text{eff}
\end{equation}
with the effective lattice momentum $\hat{k}_i^\text{eff}$ as
\begin{equation}
\Lambda_{ij,lm}(\mathbf{k})=P_{il}(\mathbf{k})P_{jm}(\mathbf{k}) - \frac{1}{2}P_{ij}(\mathbf{k})P_{lm}(\mathbf{k}).
\label{eq:Lambda_ijlm}
\end{equation}
The exact definition of $\hat{k}_i^\text{eff}$ depends on the simulation details; see section~\ref{sec:gwspectra} for further details.
Typically, the EOMs in configuration space can then be expressed as 
\begin{equation}
    \ddot{u}_{ij}+3H\dot{u}_{ij}-\frac{1}{a^2}\nabla^2 u_{ij} = 16\pi G T_{ij},
\label{eq:GWEOM}
\end{equation}
where $H$ is the Hubble rate, $a$ the scale factor, $T_{ij}$ the stress-tensor sourcing the gravitational waves, and $G$ the gravitational constant. Since the explicit form of Eq.~\ref{eq:GWEOM} is problem specific the EOMs have to be implemented for each project separately. This, however, can be done easily by adding the kernel function 

\begin{lstlisting}[language=c++]
template<> 
AMREX_GPU_DEVICE AMREX_FORCE_INLINE
void GravitationalWavesRhs<true>(
    const amrex::Array4<double>& rhs,
    const amrex::Array4<const double>& state,
    const int i, const int j, const int k,
    const int lev, const double time,
    const double dt, const double dx, 
    const double* params) {
    // Compute Rhs of EoM here ...
}
\end{lstlisting}

The procedure here is precisely the same as for the \texttt{Rhs} kernel function, except now for the auxiliary field $u$ and its conjugate momenta $\dot{u}$. The objects \texttt{rhs} and \texttt{state} will now include not only the usual scalar field components but also the auxiliary fields. Where the scalar field components can be accessed through \texttt{Scalar::\$MyScalar}, the auxiliary fields $u$ and $\dot{u}$ can be accessed with \texttt{Gw::u\_\$i\$j} and \texttt{Gw::du\_\$i\$j}, respectively. Here, \texttt{\$i\$j} is short for the 6 degrees of freedom \texttt{xx}, \texttt{yy}, \texttt{zz}, \texttt{xy}, \texttt{xz}, and \texttt{yz}. Analogously, \texttt{params} can be set by overriding the virtual function \texttt{SetParamsGravitationalWaveRhs}. The corresponding unbinned gravitational-wave spectrum can be computed automatically; see section~\ref{sec:gwspectra} for details. 

\section{Simulation Input}
\label{sec:input}
At the beginning of a simulation, all field components will be initialized to zero, though we typically want to overwrite these with some pregenerated state. To do so, we can save the initial state as a dataset in an \texttt{hdf5} file and add \texttt{input.initial\_state = \$PathToFile} to the \texttt{inputs} file. The name of the dataset should correspond to the name that was given to each scalar in the \texttt{SLEDGEHAMR\_ADD\_SCALARS} and \texttt{SLEDGEHAMR\_ADD\_CONJUGATE\_MOMENTA} macros. Alternatively, we can provide a file for each component separately through \texttt{input.initial\_state\_\$MyScalar = \$PathToFile}, in which case \texttt{data} is an acceptable dataset name as well.

For very large simulations with a coarse-level resolution of $2048^3$ and larger, it can be beneficial to split up the initial state into chunks rather than providing it in one large file. Here, our strategy is to use one chunk per compute node, so the first step is to determine the proper chunk size and location. This can be done easily by temporarily adding \texttt{input.get\_box\_layout\_nodes = \#nodes} to the \texttt{inputs} file. In the presence of this parameter, \sledgehamr\ will not start a simulation but instead write an hdf5 file containing a list of chunk locations.\footnote{In this case, \sledgehamr\ can be run on a single core as no major memory will be allocated.} This list can then be used to prepare the initial state. To feed the initial state into \sledgehamr, we can set \texttt{input.initial\_state} to the folder containing the individual chunks. Here, we need to follow the naming convention \texttt{\$MyScalar\_\$i.hdf5}, where \texttt{\$i} is the number of the chunk.

The size of the provided initial state should match that of the coarse level. However, we can provide a down-sampled version and use \texttt{input.upsample=2} to up-scale the initial state to the coarse-level resolution through interpolation. Any upsample factor that is a power of 2 is acceptable, as long as it is smaller than the blocking factor.

\section{Simulation Output}
\label{sec:output}
\sledgehamr\ provides various types of output that can be written periodically to a folder set by \texttt{output.output\_folder} in the \texttt{inputs} file. Some of the different output types are described in the following subsections, but they can all be controlled by following the same pattern. Setting \texttt{output.*.interval} to a positive floating-point value in the inputs file will allow the output to be written in regular intervals. We can also set a minimum (maximum) time before (after) which no output will be written. This is controlled by \texttt{output.*.min\_t} (\texttt{output.*.max\_t}). Here, the asterisk corresponds to the output types, which are
\begin{itemize}
    \itemsep0.2em
    \item \texttt{checkpoints}
    \item \texttt{slices}
    \item \texttt{coarse\_box}
    \item \texttt{full\_box}
    \item \texttt{slices\_truncation\_error}
    \item \texttt{coarse\_box\_truncation\_error}
    \item \texttt{full\_box\_truncation\_error}
    \item \texttt{projections}
    \item \texttt{spectra}
    \item \texttt{gw\_spectra}
    \item \texttt{performance\_monitor}
    \item \texttt{amrex\_plotfile}
\end{itemize}
 These different types are explained in the next few sections.
 
Sometimes, certain output types can take up a significant amount of disk space. To help distribute disk usage, we can define an alternative output folder through \texttt{output.alternative\_output\_folder}. If for any output type \texttt{output.*.alternate} is set to $1$ then every second output of this type will be written in the alternative output folder location rather than the primary. 

An interface to easily write problem-specific custom output is provided as well; see section \ref{sec:CustomOutputs}. Other parameters to control the output behavior can be found in appendix~\ref{sec:parameters}. We also provided the Python-based tool \texttt{pySledgehamr} to easily read and visualize the different outputs. To use \texttt{pySledgehamr} we simply need to create an instance of the \texttt{output} class:
\begin{lstlisting}[language=python]
import sys
sys.path.append(PathToSledgehamrDirectory)
import pySledgehamr as sl
output = sl.Output(PathToOutputDirectory)
\end{lstlisting}
The object \texttt{output} can then be used to load the simulation output, the details of which are explained in the following.

\subsection{Slices and Volumes}
\label{sec:slices}
There are three different output types that directly output the field state of all scalar components: \texttt{slices}, \texttt{coarse\_box}, and \texttt{full\_box}. 
\texttt{slices} will output three slices through the field at $x=0$, $y=0$, and $z=0$, respectively. If part of the slice has been refined, these higher levels will be written to disk as well. 
\texttt{coarse\_box} simply writes the entire coarse-level field state. By setting \texttt{output.coarse\_box.downsample\_factor} to any power of 2\footnote{Must be $\leq$ \texttt{amr.blocking\_factor}.} we can write a downsampled version of the coarse box to save disk space. 
Similarly, \texttt{full\_box} will write the entire field state, but unlike \texttt{coarse\_box} it will include all refinement levels. Also here, we can produce a downsampled version by setting \texttt{output.full\_box.downsample\_factor} appropriately.  

To load the result, we can use \texttt{pySledgehamr}:
\begin{lstlisting}[language=python]
output.GetSlice(i, direction, level, fields)
output.GetCoarseBox(i, fields)
output.GetFullBox(i, level, fields)
\end{lstlisting}
These functions will load the \texttt{i}-th state at refinement level \texttt{level} of all scalar fields included in the list \texttt{fields}. Furthermore, \texttt{direction}$\in$\texttt{\{"x", "y", "z"\}} specifies the orientation of the slices. Each function returns a dictionary containing the corresponding time $t$ and the state of each scalar field.

\subsection{Truncation Error Estimates}
\label{sec:TEs}
One of the main refinement criteria is TEEs described section~\ref{sec:tagging}. These can be saved using the output types \texttt{slices\_truncation\_error}, \texttt{coarse\_box\_truncation\_error}, and \texttt{full\_box\_truncation\_error}. These work completely analogously to \texttt{slices}, \texttt{coarse\_box}, and \texttt{full\_box}, including the downsampling factors. A version of the corresponding field state is included in the output as well to allow for a direct comparison. Note that the output interval for these output types can be slightly larger than \texttt{output.*.interval} as TEEs are only computed during regrids. \sledgehamr\ therefore has to wait for a regrid at the coarse level before these output types can be written.

The output can be loaded analogously to normal slices and boxes by adding \texttt{TruncationError} to the respective \texttt{pySledgehamr} function name.

\subsection{Projections}
\label{sec:projections}
Particularly useful output types are projections, which are 2D outputs where the third dimension is integrated out. That way we can visually observe features in the entire simulation volume without having to save the full 3D output. Projections will also include fine-level data by up-sampling the 2D output to the resolution of the finest level. The maximum level can be capped through \texttt{output.projections.max\_level}, however. Restricting the maximum level is recommended, as proper up-sampling to the target resolution can take a significant amount of time.

It is typically not useful to project the field components itself. Instead, projecting a derived quantity like an energy density is often more meaningful. Therefore, we always need to specify how to compute such quantity. To do so, we need to define a new kernel function using the following template:

\begin{lstlisting}[language=c++]
AMREX_FORCE_INLINE
double MyProjection(
    amrex::Array4<double const> const& state, const int i, const int j, const int k, 
    const int lev, const double time,
    const double dt, const double dx,
    const std::vector<double>& params) {
    // Return my derived quantity...
}
\end{lstlisting}

The function arguments are the same as those of the \texttt{rhs} function; see section~\ref{sec:MinimalExample}, and we need to return the derived quantity at cell $(i,j,k)$. While \texttt{params} in the \texttt{rhs} function can be set by overriding \texttt{SetParams}, we can do the equivalent here by overriding \texttt{SetParamsProjections}. To use this kernel function for a projection, all we need to do is add

\begin{lstlisting}[language=c++]
io_module->projections.emplace_back(
    MyProjection, "MyProjection");
\end{lstlisting}
to the project's \texttt{Init} function\footnote{See section~\ref{sec:virtual}.}. Here, the first argument is the name of the kernel function, whereas the second argument is a unique identifying string. Multiple different projection types can be added this way, which will all be computed simultaneously. The projections can be loaded with \texttt{pySledgehamr} using \texttt{output.GetProjection(i, ["MyProjection"])}.

\subsection{Custom Spectra}
\label{sec:spectra}
It is possible to compute unbinned 1D spectra of the form 
\begin{equation}
S_X(|\mathbf{k}|) = \frac{L t}{2\pi N^6}\sum_{\mathbf{k}=|\mathbf{k}|}|\tilde{X}(\mathbf{k})|^2,
\end{equation}
at run-time, where $L$ is the box size, $N^3$ is the total number of coarse-level cells, $t$ the current time, and $\tilde{X}$ is the Fourier transform of a quantity $X$. $X$ can be defined using the same kernel function template we use for projections; see section~\ref{sec:projections}. Adding a spectrum type to a project works entirely analogously to a projection as well by simply adding 

\begin{lstlisting}[language=c++]
io_module->spectra.emplace_back(
    MySpectrum, "MySpectrum");
\end{lstlisting}
to the project's \texttt{Init} function. All spectra will be computed using the coarse-level information to avoid the usage of a nonuniform fast Fourier transform scheme. 

Note that producing unbinned spectra involves knowing all unique combinations of $i^2+j^2+k^2$ for $i$, $j$, $k\in[0,N)$. \sledgehamr\ will load these combinations from a file to save run-time rather than recomputing them each time. The file provided by the GitHub repository contains all necessary data for coarse-level sizes of up to $N^3=1024^3$. For larger coarse levels,\footnote{They are not provided by default to avoid unnecessarily large file sizes.} the unique combinations need to be added to this file first, which can be done easily by running the notebook \texttt{notebooks/AddSpectrumBins.ipynb}. The spectra can be loaded using \texttt{output.GetSpectrum(i, ["MySpectrum"])}.

\subsection{Gravitational-Wave Spectra}
\label{sec:gwspectra}
If the simulation is evolved with gravitational waves, the corresponding spectrum can be automatically computed by setting \texttt{output.gw\_spectra.interval} to a positive value. For ultimate flexibility, the output will be the unbinned quantity 
\begin{equation}
    G(|\mathbf{k}|)=\frac{1}{N^6}\sum_{\mathbf{k}=|\mathbf{k}|}\Lambda_{ij,lm}(\mathbf{k})\tilde{\dot{u}}_{ij}(\mathbf{k})\tilde{\dot{u}}_{lm}^{*}(\mathbf{k})
    \label{eq:gw_spectrum}
\end{equation}
from which the differential spectrum $\partial \rho_\text{GW}/\partial k$ can easily be constructed following the procedure outlined in~\cite{Figueroa:2011ye}. Here, $N^3$ is the total number of coarse-level cells, $\tilde{\dot{u}}_{ij}$ is the Fourier transform of the auxilary field $\dot{u}_{ij}$ and $\Lambda_{ij,lm}$ is given in Eq.~\ref{eq:Lambda_ijlm}.
The computation of $\Lambda_{ij,lm}$ requires us to specify the effective lattice momentum $\mathbf{k}^\text{eff}$ which depends on how the gradients in the auxiliary field source term are computed. We have currently two types implemented, which can be selected by setting \texttt{output.gw\_spectra.projection\_type} to either 1 (suitable for second-order finite-difference gradients) or 2 (suitable for fourth-order finite-difference gradients). The gravitational-wave spectrum will be computed based on the coarse level. However, just like for general spectra, this requires precomputing all unique combinations of $i^2+j^2+k^2$ for $i$, $j$, $k\in[0,N)$ if the coarse-level is larger than $N^3=1024^3$. See section~\ref{sec:spectra} for details.

Furthermore, it is possible to compute gravitational-wave spectra based on a modified version of \eqref{eq:gw_spectrum}. This can be achieved by deriving a class from the base class \texttt{GravitationalWavesSpectrumModifier}, which has two virtual functions that can be overridden to change the $\tilde{u}'_{ij}(\mathbf{k})\tilde{u}_{lm}'^{*}(\mathbf{k})$ term. An instance of this class can then be passed to the routine that computes the gravitational-wave spectra, e.g.

\begin{lstlisting}[language=c++]
unique_ptr<GravitationalWavesSpectrumModifier> modifier = make_unique<MyModification>();
gravitational_waves->ComputeSpectrum(hdf5_file, modifier.get());
\end{lstlisting}
as part of a new output type\footnote{See section~\ref{sec:CustomOutputs}}. We abstain from further details here, but a full implementation example can be found in the \texttt{FirstOrderPhaseTransition} project that comes with the GitHub repository. The spectra can be loaded using \texttt{output.GetGravitationalWaveSpectrum(i)}.

\subsection{Checkpoints}
\label{sec:checkpoints}
One of the key features of any simulation code of this size are of course checkpoints. Checkpoints allow a simulation to be restarted after an early termination of the program. To periodically write checkpoints, one needs to set \texttt{output.checkpoints.interval} to a positive value. As checkpoints can take up a significant amount of disk space since they contain data of all levels at full double precision, it is often beneficial to set \texttt{output.checkpoints.rolling = 1}. In this case, only the last checkpoint is kept on disk, and any older checkpoints are deleted. To restart a simulation from a checkpoint, we need to set \texttt{input.restart = 1}. \sledgehamr\ will automatically find the latest checkpoint within the output folder, although a specific checkpoint can be selected through \texttt{input.select\_checkpoint}.

\subsection{Custom Output}
\label{sec:CustomOutputs}
In addition to the above-mentioned predefined output types, it is possible to implement extra problem-specific output. All that is needed is to add 

\begin{lstlisting}[language=c++]
io_module->output.emplace_back("MyOutput", OUTPUT_FCT(MyOutput));
\end{lstlisting}

to the project's \texttt{Init} function. Here, the first argument is a unique string to identify the output type. Through this, all typical input parameters such as \texttt{output.MyOutput.interval} become available. The second argument is a function pointer, pointing to a function of the form
\begin{lstlisting}[language=c++]
bool MyOutput(double time, std::string prefix) {
    // Write custom output here
    // ...
    return true;
}
\end{lstlisting}
Here, the actual writing of the output can be implemented, where the function arguments provide the current time and the folder \texttt{prefix} to which the output shall be written. Furthermore, we expect a boolean value as a return indicating whether the writing of the output was successful. If \texttt{false} is returned, the folder \texttt{prefix} will be deleted, and another writing attempt will be performed next interval. If \texttt{MyOutput} is a member function of the project class, we can access all level data easily through, e.g., \texttt{sim->GetLevelData(lev)}.

Besides custom output, it is also possible to modify the interval at which output is written. Output is written once $f(t_\text{Current}) > f(t_\text{LastWritten}) + \texttt{interval}$, where \texttt{interval} is set by \texttt{output.*.interval}. The default is $f(t) = t$, but this can be modified through 
\begin{lstlisting}[language=c++]
io_module->output[sim->io_module->idx_*].
    SetTimeFunction(TIME_FCT(MyF_of_t));
\end{lstlisting}
where
\begin{lstlisting}[language=c++]
double MyF_of_t(const double t) {
    return ...
}
\end{lstlisting}
Here, the asterisks in \texttt{sim->io\_module->idx\_*} shall be replaced by the output type for which we want to change $f(t)$, e.g. \texttt{sim->io\_module->idx\_checkpoints}. This also affects the minimum (maximum) time \texttt{min\_t} (\texttt{max\_t}), where we have $f(t) > \texttt{min\_t}$.

Concrete examples of custom output implementations and modifications to the write interval can be found in the \texttt{AxionStrings} project that comes with the GitHub repository.

\section{Discussion}
\label{sec:discussion}
We present \sledgehamr, a massively parallelized GPU-enabled code to simulate an arbitrary amount of coupled scalar fields on an adaptive mesh. \sledgehamr\ aims to simplify the use of AMR by balancing usability and complexity. While easy to use it provides a flexible and customizable interface that allows, for example, for the computation of gravitational-wave spectra. 

\sledgehamr\ is still in active development and will likely be expanded with additional features in the future. Older versions, however, have already been used successfully in a few publications~\cite{Buschmann:2021sdq, Benabou:2023ghl, AxionSim, FOPT}, which, for example, provide one of the most robust predictions of the QCD axion mass in a postinflationary scenario~\cite{Buschmann:2021sdq, AxionSim}.

%% Also note that the akcnowlodgment environment does not support long amounts of text. If you have a lot of people and institutions to acknowledge, do not use this command. Instead, create a new \section{Acknowledgments}.
\begin{acknowledgments}
{\it 
We would like to thank J. Benabou, J. Foster, M. Khelashvili, and A. Kunder for having used the code during various stages of its development as well as B. Safdi and W. Zhang for helpful discussions. 
We acknowledge funding from the European Research Council (ERC) under the European Union’s Horizon 2020 research and innovation programme (Grant agreement No. 864035).
This research used resources of the National Energy Research Scientific Computing Center (NERSC), a U.S. Department of Energy Office of Science User Facility located at Lawrence Berkeley National Laboratory, operated under Contract No. DE-AC02-05CH11231 using NERSC award HEP-ERCAP0023978.
}
\end{acknowledgments}

\section*{Data availability}
The code \sledgehamr\ is available on GitHub (\url{https://github.com/MSABuschmann/sledgehamr}) as well as on Zenodo~\cite{buschmann_2024_14285061}. The data
underlying this article can be reproduced using this code and will be shared on reasonable request to the corresponding author.

\bibliography{sledgehamr}{}

\begin{thebibliography}{28}%
\makeatletter
\providecommand \@ifxundefined [1]{%
 \@ifx{#1\undefined}
}%
\providecommand \@ifnum [1]{%
 \ifnum #1\expandafter \@firstoftwo
 \else \expandafter \@secondoftwo
 \fi
}%
\providecommand \@ifx [1]{%
 \ifx #1\expandafter \@firstoftwo
 \else \expandafter \@secondoftwo
 \fi
}%
\providecommand \natexlab [1]{#1}%
\providecommand \enquote  [1]{``#1''}%
\providecommand \bibnamefont  [1]{#1}%
\providecommand \bibfnamefont [1]{#1}%
\providecommand \citenamefont [1]{#1}%
\providecommand \href@noop [0]{\@secondoftwo}%
\providecommand \href [0]{\begingroup \@sanitize@url \@href}%
\providecommand \@href[1]{\@@startlink{#1}\@@href}%
\providecommand \@@href[1]{\endgroup#1\@@endlink}%
\providecommand \@sanitize@url [0]{\catcode `\\12\catcode `\$12\catcode
  `\&12\catcode `\#12\catcode `\^12\catcode `\_12\catcode `\%12\relax}%
\providecommand \@@startlink[1]{}%
\providecommand \@@endlink[0]{}%
\providecommand \url  [0]{\begingroup\@sanitize@url \@url }%
\providecommand \@url [1]{\endgroup\@href {#1}{\urlprefix }}%
\providecommand \urlprefix  [0]{URL }%
\providecommand \Eprint [0]{\href }%
\providecommand \doibase [0]{http://dx.doi.org/}%
\providecommand \selectlanguage [0]{\@gobble}%
\providecommand \bibinfo  [0]{\@secondoftwo}%
\providecommand \bibfield  [0]{\@secondoftwo}%
\providecommand \translation [1]{[#1]}%
\providecommand \BibitemOpen [0]{}%
\providecommand \bibitemStop [0]{}%
\providecommand \bibitemNoStop [0]{.\EOS\space}%
\providecommand \EOS [0]{\spacefactor3000\relax}%
\providecommand \BibitemShut  [1]{\csname bibitem#1\endcsname}%
\let\auto@bib@innerbib\@empty
%</preamble>
\bibitem [{\citenamefont {Cutting}\ \emph {et~al.}(2018)\citenamefont
  {Cutting}, \citenamefont {Hindmarsh},\ and\ \citenamefont
  {Weir}}]{Cutting:2018tjt}%
  \BibitemOpen
  \bibfield  {author} {\bibinfo {author} {\bibfnamefont {Daniel}\ \bibnamefont
  {Cutting}}, \bibinfo {author} {\bibfnamefont {Mark}\ \bibnamefont
  {Hindmarsh}}, \ and\ \bibinfo {author} {\bibfnamefont {David~J.}\
  \bibnamefont {Weir}},\ }\bibfield  {title} {\enquote {\bibinfo {title}
  {{Gravitational waves from vacuum first-order phase transitions: from the
  envelope to the lattice}},}\ }\href {\doibase 10.1103/PhysRevD.97.123513}
  {\bibfield  {journal} {\bibinfo  {journal} {Phys. Rev. D}\ }\textbf {\bibinfo
  {volume} {97}},\ \bibinfo {pages} {123513} (\bibinfo {year} {2018})},\
  \Eprint {http://arxiv.org/abs/1802.05712} {arXiv:1802.05712 [astro-ph.CO]}
  \BibitemShut {NoStop}%
\bibitem [{\citenamefont {Cutting}\ \emph {et~al.}(2021)\citenamefont
  {Cutting}, \citenamefont {Escartin}, \citenamefont {Hindmarsh},\ and\
  \citenamefont {Weir}}]{Cutting:2020nla}%
  \BibitemOpen
  \bibfield  {author} {\bibinfo {author} {\bibfnamefont {Daniel}\ \bibnamefont
  {Cutting}}, \bibinfo {author} {\bibfnamefont {Elba~Granados}\ \bibnamefont
  {Escartin}}, \bibinfo {author} {\bibfnamefont {Mark}\ \bibnamefont
  {Hindmarsh}}, \ and\ \bibinfo {author} {\bibfnamefont {David~J.}\
  \bibnamefont {Weir}},\ }\bibfield  {title} {\enquote {\bibinfo {title}
  {{Gravitational waves from vacuum first order phase transitions II: from thin
  to thick walls}},}\ }\href {\doibase 10.1103/PhysRevD.103.023531} {\bibfield
  {journal} {\bibinfo  {journal} {Phys. Rev. D}\ }\textbf {\bibinfo {volume}
  {103}},\ \bibinfo {pages} {023531} (\bibinfo {year} {2021})},\ \Eprint
  {http://arxiv.org/abs/2005.13537} {arXiv:2005.13537 [astro-ph.CO]}
  \BibitemShut {NoStop}%
\bibitem [{\citenamefont {Buschmann}\ \emph {et~al.}(2024)\citenamefont
  {Buschmann}, \citenamefont {Foster}, \citenamefont {Geller},\ and\
  \citenamefont {Opferkuch}}]{FOPT}%
  \BibitemOpen
  \bibfield  {author} {\bibinfo {author} {\bibfnamefont {Malte}\ \bibnamefont
  {Buschmann}}, \bibinfo {author} {\bibfnamefont {Joshua~W.}\ \bibnamefont
  {Foster}}, \bibinfo {author} {\bibfnamefont {Sarah~R.}\ \bibnamefont
  {Geller}}, \ and\ \bibinfo {author} {\bibfnamefont {Toby}\ \bibnamefont
  {Opferkuch}},\ }\bibfield  {title} {\enquote {\bibinfo {title} {{Thick and
  Thin Wall Collisions with Adaptive Mesh Refinement}},}\ }\href@noop {}
  {\bibfield  {journal} {\bibinfo  {journal} {to appear}\ } (\bibinfo {year}
  {2024})}\BibitemShut {NoStop}%
\bibitem [{\citenamefont {Gorghetto}\ \emph {et~al.}(2018)\citenamefont
  {Gorghetto}, \citenamefont {Hardy},\ and\ \citenamefont
  {Villadoro}}]{Gorghetto:2018myk}%
  \BibitemOpen
  \bibfield  {author} {\bibinfo {author} {\bibfnamefont {Marco}\ \bibnamefont
  {Gorghetto}}, \bibinfo {author} {\bibfnamefont {Edward}\ \bibnamefont
  {Hardy}}, \ and\ \bibinfo {author} {\bibfnamefont {Giovanni}\ \bibnamefont
  {Villadoro}},\ }\bibfield  {title} {\enquote {\bibinfo {title} {{Axions from
  Strings: the Attractive Solution}},}\ }\href {\doibase
  10.1007/JHEP07(2018)151} {\bibfield  {journal} {\bibinfo  {journal} {JHEP}\
  }\textbf {\bibinfo {volume} {07}},\ \bibinfo {pages} {151} (\bibinfo {year}
  {2018})},\ \Eprint {http://arxiv.org/abs/1806.04677} {arXiv:1806.04677
  [hep-ph]} \BibitemShut {NoStop}%
\bibitem [{\citenamefont {Gorghetto}\ \emph
  {et~al.}(2021{\natexlab{a}})\citenamefont {Gorghetto}, \citenamefont
  {Hardy},\ and\ \citenamefont {Villadoro}}]{Gorghetto:2020qws}%
  \BibitemOpen
  \bibfield  {author} {\bibinfo {author} {\bibfnamefont {Marco}\ \bibnamefont
  {Gorghetto}}, \bibinfo {author} {\bibfnamefont {Edward}\ \bibnamefont
  {Hardy}}, \ and\ \bibinfo {author} {\bibfnamefont {Giovanni}\ \bibnamefont
  {Villadoro}},\ }\bibfield  {title} {\enquote {\bibinfo {title} {{More Axions
  from Strings}},}\ }\href {\doibase 10.21468/SciPostPhys.10.2.050} {\bibfield
  {journal} {\bibinfo  {journal} {SciPost Phys.}\ }\textbf {\bibinfo {volume}
  {10}},\ \bibinfo {pages} {050} (\bibinfo {year} {2021}{\natexlab{a}})},\
  \Eprint {http://arxiv.org/abs/2007.04990} {arXiv:2007.04990 [hep-ph]}
  \BibitemShut {NoStop}%
\bibitem [{\citenamefont {Gorghetto}\ \emph
  {et~al.}(2021{\natexlab{b}})\citenamefont {Gorghetto}, \citenamefont
  {Hardy},\ and\ \citenamefont {Nicolaescu}}]{Gorghetto:2021fsn}%
  \BibitemOpen
  \bibfield  {author} {\bibinfo {author} {\bibfnamefont {Marco}\ \bibnamefont
  {Gorghetto}}, \bibinfo {author} {\bibfnamefont {Edward}\ \bibnamefont
  {Hardy}}, \ and\ \bibinfo {author} {\bibfnamefont {Horia}\ \bibnamefont
  {Nicolaescu}},\ }\bibfield  {title} {\enquote {\bibinfo {title} {{Observing
  invisible axions with gravitational waves}},}\ }\href {\doibase
  10.1088/1475-7516/2021/06/034} {\bibfield  {journal} {\bibinfo  {journal}
  {JCAP}\ }\textbf {\bibinfo {volume} {06}},\ \bibinfo {pages} {034} (\bibinfo
  {year} {2021}{\natexlab{b}})},\ \Eprint {http://arxiv.org/abs/2101.11007}
  {arXiv:2101.11007 [hep-ph]} \BibitemShut {NoStop}%
\bibitem [{\citenamefont {Buschmann}\ \emph {et~al.}(2022)\citenamefont
  {Buschmann}, \citenamefont {Foster}, \citenamefont {Hook}, \citenamefont
  {Peterson}, \citenamefont {Willcox}, \citenamefont {Zhang},\ and\
  \citenamefont {Safdi}}]{Buschmann:2021sdq}%
  \BibitemOpen
  \bibfield  {author} {\bibinfo {author} {\bibfnamefont {Malte}\ \bibnamefont
  {Buschmann}}, \bibinfo {author} {\bibfnamefont {Joshua~W.}\ \bibnamefont
  {Foster}}, \bibinfo {author} {\bibfnamefont {Anson}\ \bibnamefont {Hook}},
  \bibinfo {author} {\bibfnamefont {Adam}\ \bibnamefont {Peterson}}, \bibinfo
  {author} {\bibfnamefont {Don~E.}\ \bibnamefont {Willcox}}, \bibinfo {author}
  {\bibfnamefont {Weiqun}\ \bibnamefont {Zhang}}, \ and\ \bibinfo {author}
  {\bibfnamefont {Benjamin~R.}\ \bibnamefont {Safdi}},\ }\bibfield  {title}
  {\enquote {\bibinfo {title} {{Dark matter from axion strings with adaptive
  mesh refinement}},}\ }\href {\doibase 10.1038/s41467-022-28669-y} {\bibfield
  {journal} {\bibinfo  {journal} {Nature Commun.}\ }\textbf {\bibinfo {volume}
  {13}},\ \bibinfo {pages} {1049} (\bibinfo {year} {2022})},\ \Eprint
  {http://arxiv.org/abs/2108.05368} {arXiv:2108.05368 [hep-ph]} \BibitemShut
  {NoStop}%
\bibitem [{\citenamefont {Benabou}\ \emph
  {et~al.}(2024{\natexlab{a}})\citenamefont {Benabou}, \citenamefont
  {Buschmann}, \citenamefont {Kumar}, \citenamefont {Park},\ and\ \citenamefont
  {Safdi}}]{Benabou:2023ghl}%
  \BibitemOpen
  \bibfield  {author} {\bibinfo {author} {\bibfnamefont {Joshua~N.}\
  \bibnamefont {Benabou}}, \bibinfo {author} {\bibfnamefont {Malte}\
  \bibnamefont {Buschmann}}, \bibinfo {author} {\bibfnamefont {Soubhik}\
  \bibnamefont {Kumar}}, \bibinfo {author} {\bibfnamefont {Yujin}\ \bibnamefont
  {Park}}, \ and\ \bibinfo {author} {\bibfnamefont {Benjamin~R.}\ \bibnamefont
  {Safdi}},\ }\bibfield  {title} {\enquote {\bibinfo {title} {{Signatures of
  primordial energy injection from axion strings}},}\ }\href {\doibase
  10.1103/PhysRevD.109.055005} {\bibfield  {journal} {\bibinfo  {journal}
  {Phys. Rev. D}\ }\textbf {\bibinfo {volume} {109}},\ \bibinfo {pages}
  {055005} (\bibinfo {year} {2024}{\natexlab{a}})},\ \Eprint
  {http://arxiv.org/abs/2308.01334} {arXiv:2308.01334 [hep-ph]} \BibitemShut
  {NoStop}%
\bibitem [{\citenamefont {Saikawa}\ \emph {et~al.}(2024)\citenamefont
  {Saikawa}, \citenamefont {Redondo}, \citenamefont {Vaquero},\ and\
  \citenamefont {Kaltschmidt}}]{Saikawa:2024bta}%
  \BibitemOpen
  \bibfield  {author} {\bibinfo {author} {\bibfnamefont {Ken'ichi}\
  \bibnamefont {Saikawa}}, \bibinfo {author} {\bibfnamefont {Javier}\
  \bibnamefont {Redondo}}, \bibinfo {author} {\bibfnamefont {Alejandro}\
  \bibnamefont {Vaquero}}, \ and\ \bibinfo {author} {\bibfnamefont {Mathieu}\
  \bibnamefont {Kaltschmidt}},\ }\bibfield  {title} {\enquote {\bibinfo {title}
  {{Spectrum of global string networks and the axion dark matter mass}},}\
  }\href@noop {} {\  (\bibinfo {year} {2024})},\ \Eprint
  {http://arxiv.org/abs/2401.17253} {arXiv:2401.17253 [hep-ph]} \BibitemShut
  {NoStop}%
\bibitem [{\citenamefont {Kim}\ \emph {et~al.}(2024)\citenamefont {Kim},
  \citenamefont {Park},\ and\ \citenamefont {Son}}]{Kim:2024wku}%
  \BibitemOpen
  \bibfield  {author} {\bibinfo {author} {\bibfnamefont {Heejoo}\ \bibnamefont
  {Kim}}, \bibinfo {author} {\bibfnamefont {Junghyeon}\ \bibnamefont {Park}}, \
  and\ \bibinfo {author} {\bibfnamefont {Minho}\ \bibnamefont {Son}},\
  }\bibfield  {title} {\enquote {\bibinfo {title} {{Axion Dark Matter from
  Cosmic String Network}},}\ }\href@noop {} {\  (\bibinfo {year} {2024})},\
  \Eprint {http://arxiv.org/abs/2402.00741} {arXiv:2402.00741 [hep-ph]}
  \BibitemShut {NoStop}%
\bibitem [{\citenamefont {Benabou}\ \emph
  {et~al.}(2024{\natexlab{b}})\citenamefont {Benabou}, \citenamefont
  {Buschmann}, \citenamefont {Foster},\ and\ \citenamefont {Safdi}}]{AxionSim}%
  \BibitemOpen
  \bibfield  {author} {\bibinfo {author} {\bibfnamefont {Joshua~N.}\
  \bibnamefont {Benabou}}, \bibinfo {author} {\bibfnamefont {Malte}\
  \bibnamefont {Buschmann}}, \bibinfo {author} {\bibfnamefont {Joshua~W.}\
  \bibnamefont {Foster}}, \ and\ \bibinfo {author} {\bibfnamefont
  {Benjamin~R.}\ \bibnamefont {Safdi}},\ }\bibfield  {title} {\enquote
  {\bibinfo {title} {{Axion mass prediction from adaptive mesh refinement
  cosmological lattice simulations}},}\ }\href@noop {} {\  (\bibinfo {year}
  {2024}{\natexlab{b}})},\ \Eprint {http://arxiv.org/abs/2412.08699}
  {arXiv:2412.08699 [hep-ph]} \BibitemShut {NoStop}%
\bibitem [{\citenamefont {Berger}\ and\ \citenamefont
  {Oliger}(1984)}]{Berger:1984zza}%
  \BibitemOpen
  \bibfield  {author} {\bibinfo {author} {\bibfnamefont {Marsha~J.}\
  \bibnamefont {Berger}}\ and\ \bibinfo {author} {\bibfnamefont {Joseph}\
  \bibnamefont {Oliger}},\ }\bibfield  {title} {\enquote {\bibinfo {title}
  {{Adaptive Mesh Refinement for Hyperbolic Partial Differential Equations}},}\
  }\href {\doibase 10.1016/0021-9991(84)90073-1} {\bibfield  {journal}
  {\bibinfo  {journal} {J. Comput. Phys.}\ }\textbf {\bibinfo {volume} {53}},\
  \bibinfo {pages} {484} (\bibinfo {year} {1984})}\BibitemShut {NoStop}%
\bibitem [{\citenamefont {Berger}\ and\ \citenamefont
  {Colella}(1989)}]{BERGER198964}%
  \BibitemOpen
  \bibfield  {author} {\bibinfo {author} {\bibfnamefont {M.J.}\ \bibnamefont
  {Berger}}\ and\ \bibinfo {author} {\bibfnamefont {P.}~\bibnamefont
  {Colella}},\ }\bibfield  {title} {\enquote {\bibinfo {title} {Local adaptive
  mesh refinement for shock hydrodynamics},}\ }\href {\doibase
  https://doi.org/10.1016/0021-9991(89)90035-1} {\bibfield  {journal} {\bibinfo
   {journal} {Journal of Computational Physics}\ }\textbf {\bibinfo {volume}
  {82}},\ \bibinfo {pages} {64--84} (\bibinfo {year} {1989})}\BibitemShut
  {NoStop}%
\bibitem [{\citenamefont {Bell}\ \emph {et~al.}(1994)\citenamefont {Bell},
  \citenamefont {Berger}, \citenamefont {Saltzman},\ and\ \citenamefont
  {Welcome}}]{doi:10.1137/0915008}%
  \BibitemOpen
  \bibfield  {author} {\bibinfo {author} {\bibfnamefont {John}\ \bibnamefont
  {Bell}}, \bibinfo {author} {\bibfnamefont {Marsha}\ \bibnamefont {Berger}},
  \bibinfo {author} {\bibfnamefont {Jeff}\ \bibnamefont {Saltzman}}, \ and\
  \bibinfo {author} {\bibfnamefont {Mike}\ \bibnamefont {Welcome}},\ }\bibfield
   {title} {\enquote {\bibinfo {title} {Three-dimensional adaptive mesh
  refinement for hyperbolic conservation laws},}\ }\href {\doibase
  10.1137/0915008} {\bibfield  {journal} {\bibinfo  {journal} {SIAM Journal on
  Scientific Computing}\ }\textbf {\bibinfo {volume} {15}},\ \bibinfo {pages}
  {127--138} (\bibinfo {year} {1994})},\ \Eprint
  {http://arxiv.org/abs/https://doi.org/10.1137/0915008}
  {https://doi.org/10.1137/0915008} \BibitemShut {NoStop}%
\bibitem [{\citenamefont {Bryan}\ \emph {et~al.}(2014)\citenamefont {Bryan},
  \citenamefont {Norman}, \citenamefont {O’Shea}, \citenamefont {Abel},
  \citenamefont {Wise}, \citenamefont {Turk}, \citenamefont {Reynolds},
  \citenamefont {Collins}, \citenamefont {Wang}, \citenamefont {Skillman},
  \citenamefont {Smith}, \citenamefont {Harkness}, \citenamefont {Bordner},
  \citenamefont {Kim}, \citenamefont {Kuhlen}, \citenamefont {Xu},
  \citenamefont {Goldbaum}, \citenamefont {Hummels}, \citenamefont {Kritsuk},
  \citenamefont {Tasker}, \citenamefont {Skory}, \citenamefont {Simpson},
  \citenamefont {Hahn}, \citenamefont {Oishi}, \citenamefont {So},
  \citenamefont {Zhao}, \citenamefont {Cen},\ and\ \citenamefont
  {Li}}]{Bryan_2014}%
  \BibitemOpen
  \bibfield  {author} {\bibinfo {author} {\bibfnamefont {Greg~L.}\ \bibnamefont
  {Bryan}}, \bibinfo {author} {\bibfnamefont {Michael~L.}\ \bibnamefont
  {Norman}}, \bibinfo {author} {\bibfnamefont {Brian~W.}\ \bibnamefont
  {O’Shea}}, \bibinfo {author} {\bibfnamefont {Tom}\ \bibnamefont {Abel}},
  \bibinfo {author} {\bibfnamefont {John~H.}\ \bibnamefont {Wise}}, \bibinfo
  {author} {\bibfnamefont {Matthew~J.}\ \bibnamefont {Turk}}, \bibinfo {author}
  {\bibfnamefont {Daniel~R.}\ \bibnamefont {Reynolds}}, \bibinfo {author}
  {\bibfnamefont {David~C.}\ \bibnamefont {Collins}}, \bibinfo {author}
  {\bibfnamefont {Peng}\ \bibnamefont {Wang}}, \bibinfo {author} {\bibfnamefont
  {Samuel~W.}\ \bibnamefont {Skillman}}, \bibinfo {author} {\bibfnamefont
  {Britton}\ \bibnamefont {Smith}}, \bibinfo {author} {\bibfnamefont
  {Robert~P.}\ \bibnamefont {Harkness}}, \bibinfo {author} {\bibfnamefont
  {James}\ \bibnamefont {Bordner}}, \bibinfo {author} {\bibfnamefont {Ji-hoon}\
  \bibnamefont {Kim}}, \bibinfo {author} {\bibfnamefont {Michael}\ \bibnamefont
  {Kuhlen}}, \bibinfo {author} {\bibfnamefont {Hao}\ \bibnamefont {Xu}},
  \bibinfo {author} {\bibfnamefont {Nathan}\ \bibnamefont {Goldbaum}}, \bibinfo
  {author} {\bibfnamefont {Cameron}\ \bibnamefont {Hummels}}, \bibinfo {author}
  {\bibfnamefont {Alexei~G.}\ \bibnamefont {Kritsuk}}, \bibinfo {author}
  {\bibfnamefont {Elizabeth}\ \bibnamefont {Tasker}}, \bibinfo {author}
  {\bibfnamefont {Stephen}\ \bibnamefont {Skory}}, \bibinfo {author}
  {\bibfnamefont {Christine~M.}\ \bibnamefont {Simpson}}, \bibinfo {author}
  {\bibfnamefont {Oliver}\ \bibnamefont {Hahn}}, \bibinfo {author}
  {\bibfnamefont {Jeffrey~S.}\ \bibnamefont {Oishi}}, \bibinfo {author}
  {\bibfnamefont {Geoffrey~C.}\ \bibnamefont {So}}, \bibinfo {author}
  {\bibfnamefont {Fen}\ \bibnamefont {Zhao}}, \bibinfo {author} {\bibfnamefont
  {Renyue}\ \bibnamefont {Cen}}, \ and\ \bibinfo {author} {\bibfnamefont
  {Yuan}\ \bibnamefont {Li}},\ }\bibfield  {title} {\enquote {\bibinfo {title}
  {Enzo: An adaptive mesh refinement code for astrophysics},}\ }\href {\doibase
  10.1088/0067-0049/211/2/19} {\bibfield  {journal} {\bibinfo  {journal} {The
  Astrophysical Journal Supplement Series}\ }\textbf {\bibinfo {volume}
  {211}},\ \bibinfo {pages} {19} (\bibinfo {year} {2014})}\BibitemShut
  {NoStop}%
\bibitem [{\citenamefont {Teyssier}(2002)}]{Teyssier:2001cp}%
  \BibitemOpen
  \bibfield  {author} {\bibinfo {author} {\bibfnamefont {Romain}\ \bibnamefont
  {Teyssier}},\ }\bibfield  {title} {\enquote {\bibinfo {title} {{Cosmological
  hydrodynamics with adaptive mesh refinement: a new high resolution code
  called RAMSES}},}\ }\href {\doibase 10.1051/0004-6361:20011817} {\bibfield
  {journal} {\bibinfo  {journal} {Astron. Astrophys.}\ }\textbf {\bibinfo
  {volume} {385}},\ \bibinfo {pages} {337--364} (\bibinfo {year} {2002})},\
  \Eprint {http://arxiv.org/abs/astro-ph/0111367} {arXiv:astro-ph/0111367}
  \BibitemShut {NoStop}%
\bibitem [{\citenamefont {Fryxell}\ \emph {et~al.}(2000)\citenamefont
  {Fryxell}, \citenamefont {Olson}, \citenamefont {Ricker}, \citenamefont
  {Timmes}, \citenamefont {Zingale}, \citenamefont {Lamb}, \citenamefont
  {MacNeice}, \citenamefont {Rosner}, \citenamefont {Truran},\ and\
  \citenamefont {Tufo}}]{Fryxell_2000}%
  \BibitemOpen
  \bibfield  {author} {\bibinfo {author} {\bibfnamefont {B.}~\bibnamefont
  {Fryxell}}, \bibinfo {author} {\bibfnamefont {K.}~\bibnamefont {Olson}},
  \bibinfo {author} {\bibfnamefont {P.}~\bibnamefont {Ricker}}, \bibinfo
  {author} {\bibfnamefont {F.~X.}\ \bibnamefont {Timmes}}, \bibinfo {author}
  {\bibfnamefont {M.}~\bibnamefont {Zingale}}, \bibinfo {author} {\bibfnamefont
  {D.~Q.}\ \bibnamefont {Lamb}}, \bibinfo {author} {\bibfnamefont
  {P.}~\bibnamefont {MacNeice}}, \bibinfo {author} {\bibfnamefont
  {R.}~\bibnamefont {Rosner}}, \bibinfo {author} {\bibfnamefont {J.~W.}\
  \bibnamefont {Truran}}, \ and\ \bibinfo {author} {\bibfnamefont
  {H.}~\bibnamefont {Tufo}},\ }\bibfield  {title} {\enquote {\bibinfo {title}
  {Flash: An adaptive mesh hydrodynamics code for modeling astrophysical
  thermonuclear flashes},}\ }\href {\doibase 10.1086/317361} {\bibfield
  {journal} {\bibinfo  {journal} {The Astrophysical Journal Supplement Series}\
  }\textbf {\bibinfo {volume} {131}},\ \bibinfo {pages} {273} (\bibinfo {year}
  {2000})}\BibitemShut {NoStop}%
\bibitem [{\citenamefont {Fan}\ \emph {et~al.}(2019)\citenamefont {Fan},
  \citenamefont {Nonaka}, \citenamefont {Almgren}, \citenamefont {Harpole},\
  and\ \citenamefont {Zingale}}]{Fan_2019}%
  \BibitemOpen
  \bibfield  {author} {\bibinfo {author} {\bibfnamefont {Duoming}\ \bibnamefont
  {Fan}}, \bibinfo {author} {\bibfnamefont {Andrew}\ \bibnamefont {Nonaka}},
  \bibinfo {author} {\bibfnamefont {Ann~S.}\ \bibnamefont {Almgren}}, \bibinfo
  {author} {\bibfnamefont {Alice}\ \bibnamefont {Harpole}}, \ and\ \bibinfo
  {author} {\bibfnamefont {Michael}\ \bibnamefont {Zingale}},\ }\bibfield
  {title} {\enquote {\bibinfo {title} {Maestroex: A massively parallel low mach
  number astrophysical solver},}\ }\href {\doibase 10.3847/1538-4357/ab4f75}
  {\bibfield  {journal} {\bibinfo  {journal} {The Astrophysical Journal}\
  }\textbf {\bibinfo {volume} {887}},\ \bibinfo {pages} {212} (\bibinfo {year}
  {2019})}\BibitemShut {NoStop}%
\bibitem [{\citenamefont {Clough}\ \emph {et~al.}(2015)\citenamefont {Clough},
  \citenamefont {Figueras}, \citenamefont {Finkel}, \citenamefont {Kunesch},
  \citenamefont {Lim},\ and\ \citenamefont {Tunyasuvunakool}}]{Clough:2015sqa}%
  \BibitemOpen
  \bibfield  {author} {\bibinfo {author} {\bibfnamefont {Katy}\ \bibnamefont
  {Clough}}, \bibinfo {author} {\bibfnamefont {Pau}\ \bibnamefont {Figueras}},
  \bibinfo {author} {\bibfnamefont {Hal}\ \bibnamefont {Finkel}}, \bibinfo
  {author} {\bibfnamefont {Markus}\ \bibnamefont {Kunesch}}, \bibinfo {author}
  {\bibfnamefont {Eugene~A.}\ \bibnamefont {Lim}}, \ and\ \bibinfo {author}
  {\bibfnamefont {Saran}\ \bibnamefont {Tunyasuvunakool}},\ }\bibfield  {title}
  {\enquote {\bibinfo {title} {{GRChombo : Numerical Relativity with Adaptive
  Mesh Refinement}},}\ }\href {\doibase 10.1088/0264-9381/32/24/245011}
  {\bibfield  {journal} {\bibinfo  {journal} {Class. Quant. Grav.}\ }\textbf
  {\bibinfo {volume} {32}},\ \bibinfo {pages} {245011} (\bibinfo {year}
  {2015})},\ \Eprint {http://arxiv.org/abs/1503.03436} {arXiv:1503.03436
  [gr-qc]} \BibitemShut {NoStop}%
\bibitem [{\citenamefont {Zhang}\ \emph {et~al.}(2019)\citenamefont {Zhang},
  \citenamefont {Almgren}, \citenamefont {Beckner}, \citenamefont {Bell},
  \citenamefont {Blaschke}, \citenamefont {Chan}, \citenamefont {Day},
  \citenamefont {Friesen}, \citenamefont {Gott}, \citenamefont {Graves},
  \citenamefont {Katz}, \citenamefont {Myers}, \citenamefont {Nguyen},
  \citenamefont {Nonaka}, \citenamefont {Rosso}, \citenamefont {Williams},\
  and\ \citenamefont {Zingale}}]{AMReX_JOSS}%
  \BibitemOpen
  \bibfield  {author} {\bibinfo {author} {\bibfnamefont {Weiqun}\ \bibnamefont
  {Zhang}}, \bibinfo {author} {\bibfnamefont {Ann}\ \bibnamefont {Almgren}},
  \bibinfo {author} {\bibfnamefont {Vince}\ \bibnamefont {Beckner}}, \bibinfo
  {author} {\bibfnamefont {John}\ \bibnamefont {Bell}}, \bibinfo {author}
  {\bibfnamefont {Johannes}\ \bibnamefont {Blaschke}}, \bibinfo {author}
  {\bibfnamefont {Cy}~\bibnamefont {Chan}}, \bibinfo {author} {\bibfnamefont
  {Marcus}\ \bibnamefont {Day}}, \bibinfo {author} {\bibfnamefont {Brian}\
  \bibnamefont {Friesen}}, \bibinfo {author} {\bibfnamefont {Kevin}\
  \bibnamefont {Gott}}, \bibinfo {author} {\bibfnamefont {Daniel}\ \bibnamefont
  {Graves}}, \bibinfo {author} {\bibfnamefont {Max}\ \bibnamefont {Katz}},
  \bibinfo {author} {\bibfnamefont {Andrew}\ \bibnamefont {Myers}}, \bibinfo
  {author} {\bibfnamefont {Tan}\ \bibnamefont {Nguyen}}, \bibinfo {author}
  {\bibfnamefont {Andrew}\ \bibnamefont {Nonaka}}, \bibinfo {author}
  {\bibfnamefont {Michele}\ \bibnamefont {Rosso}}, \bibinfo {author}
  {\bibfnamefont {Samuel}\ \bibnamefont {Williams}}, \ and\ \bibinfo {author}
  {\bibfnamefont {Michael}\ \bibnamefont {Zingale}},\ }\bibfield  {title}
  {\enquote {\bibinfo {title} {{AMReX}: a framework for block-structured
  adaptive mesh refinement},}\ }\href {\doibase 10.21105/joss.01370} {\bibfield
   {journal} {\bibinfo  {journal} {Journal of Open Source Software}\ }\textbf
  {\bibinfo {volume} {4}},\ \bibinfo {pages} {1370} (\bibinfo {year}
  {2019})}\BibitemShut {NoStop}%
\bibitem [{\citenamefont {Courant}\ \emph {et~al.}(1928)\citenamefont
  {Courant}, \citenamefont {Friedrichs},\ and\ \citenamefont
  {Lewy}}]{Courant:1928lij}%
  \BibitemOpen
  \bibfield  {author} {\bibinfo {author} {\bibfnamefont {R.}~\bibnamefont
  {Courant}}, \bibinfo {author} {\bibfnamefont {K.}~\bibnamefont {Friedrichs}},
  \ and\ \bibinfo {author} {\bibfnamefont {H.}~\bibnamefont {Lewy}},\
  }\bibfield  {title} {\enquote {\bibinfo {title} {{\"Uber die partiellen
  Differenzengleichungen der mathematischen Physik}},}\ }\href {\doibase
  10.1007/BF01448839} {\bibfield  {journal} {\bibinfo  {journal} {Math. Ann.}\
  }\textbf {\bibinfo {volume} {100}},\ \bibinfo {pages} {32--74} (\bibinfo
  {year} {1928})}\BibitemShut {NoStop}%
\bibitem [{\citenamefont {Almgren}\ \emph {et~al.}(2010)\citenamefont
  {Almgren}, \citenamefont {Beckner}, \citenamefont {Bell}, \citenamefont
  {Day}, \citenamefont {Howell}, \citenamefont {Joggerst}, \citenamefont
  {Lijewski}, \citenamefont {Nonaka}, \citenamefont {Singer},\ and\
  \citenamefont {Zingale}}]{Almgren_2010}%
  \BibitemOpen
  \bibfield  {author} {\bibinfo {author} {\bibfnamefont {A.~S.}\ \bibnamefont
  {Almgren}}, \bibinfo {author} {\bibfnamefont {V.~E.}\ \bibnamefont
  {Beckner}}, \bibinfo {author} {\bibfnamefont {J.~B.}\ \bibnamefont {Bell}},
  \bibinfo {author} {\bibfnamefont {M.~S.}\ \bibnamefont {Day}}, \bibinfo
  {author} {\bibfnamefont {L.~H.}\ \bibnamefont {Howell}}, \bibinfo {author}
  {\bibfnamefont {C.~C.}\ \bibnamefont {Joggerst}}, \bibinfo {author}
  {\bibfnamefont {M.~J.}\ \bibnamefont {Lijewski}}, \bibinfo {author}
  {\bibfnamefont {A.}~\bibnamefont {Nonaka}}, \bibinfo {author} {\bibfnamefont
  {M.}~\bibnamefont {Singer}}, \ and\ \bibinfo {author} {\bibfnamefont
  {M.}~\bibnamefont {Zingale}},\ }\bibfield  {title} {\enquote {\bibinfo
  {title} {Castro: A new compressible astrophysical solver. i. hydrodynamics
  and self-gravity},}\ }\href {\doibase 10.1088/0004-637x/715/2/1221}
  {\bibfield  {journal} {\bibinfo  {journal} {The Astrophysical Journal}\
  }\textbf {\bibinfo {volume} {715}},\ \bibinfo {pages} {1221–1238} (\bibinfo
  {year} {2010})}\BibitemShut {NoStop}%
\bibitem [{\citenamefont {Berger}\ and\ \citenamefont
  {Rigoutsos}(1991)}]{120081}%
  \BibitemOpen
  \bibfield  {author} {\bibinfo {author} {\bibfnamefont {M.}~\bibnamefont
  {Berger}}\ and\ \bibinfo {author} {\bibfnamefont {I.}~\bibnamefont
  {Rigoutsos}},\ }\bibfield  {title} {\enquote {\bibinfo {title} {An algorithm
  for point clustering and grid generation},}\ }\href {\doibase
  10.1109/21.120081} {\bibfield  {journal} {\bibinfo  {journal} {IEEE
  Transactions on Systems, Man, and Cybernetics}\ }\textbf {\bibinfo {volume}
  {21}},\ \bibinfo {pages} {1278--1286} (\bibinfo {year} {1991})}\BibitemShut
  {NoStop}%
\bibitem [{\citenamefont {Li}(2010)}]{LI20103139}%
  \BibitemOpen
  \bibfield  {author} {\bibinfo {author} {\bibfnamefont {Shengtai}\
  \bibnamefont {Li}},\ }\bibfield  {title} {\enquote {\bibinfo {title}
  {Comparison of refinement criteria for structured adaptive mesh
  refinement},}\ }\href {\doibase https://doi.org/10.1016/j.cam.2009.08.104}
  {\bibfield  {journal} {\bibinfo  {journal} {Journal of Computational and
  Applied Mathematics}\ }\textbf {\bibinfo {volume} {233}},\ \bibinfo {pages}
  {3139--3147} (\bibinfo {year} {2010})},\ \bibinfo {note} {finite Element
  Methods in Engineering and Science (FEMTEC 2009)}\BibitemShut {NoStop}%
\bibitem [{\citenamefont {Gustafsson}\ \emph {et~al.}(2013)\citenamefont
  {Gustafsson}, \citenamefont {Kreiss},\ and\ \citenamefont
  {Oliger}}]{KOdissipation}%
  \BibitemOpen
  \bibfield  {author} {\bibinfo {author} {\bibfnamefont {B.}~\bibnamefont
  {Gustafsson}}, \bibinfo {author} {\bibfnamefont {H-O}\ \bibnamefont
  {Kreiss}}, \ and\ \bibinfo {author} {\bibfnamefont {J.}~\bibnamefont
  {Oliger}},\ }\enquote {\bibinfo {title} {Time‐dependent problems and
  difference methods},}\ in\ \href {\doibase
  https://doi.org/10.1002/9781118548448.fmatter} {\emph {\bibinfo {booktitle}
  {Time‐Dependent Problems and Difference Methods}}}\ (\bibinfo  {publisher}
  {John Wiley \& Sons, Ltd},\ \bibinfo {year} {2013})\ \Eprint
  {http://arxiv.org/abs/https://onlinelibrary.wiley.com/doi/pdf/10.1002/9781118548448.fmatter}
  {https://onlinelibrary.wiley.com/doi/pdf/10.1002/9781118548448.fmatter}
  \BibitemShut {NoStop}%
\bibitem [{\citenamefont {Figueroa}\ \emph {et~al.}(2011)\citenamefont
  {Figueroa}, \citenamefont {Garcia-Bellido},\ and\ \citenamefont
  {Rajantie}}]{Figueroa:2011ye}%
  \BibitemOpen
  \bibfield  {author} {\bibinfo {author} {\bibfnamefont {Daniel~G.}\
  \bibnamefont {Figueroa}}, \bibinfo {author} {\bibfnamefont {Juan}\
  \bibnamefont {Garcia-Bellido}}, \ and\ \bibinfo {author} {\bibfnamefont
  {Arttu}\ \bibnamefont {Rajantie}},\ }\bibfield  {title} {\enquote {\bibinfo
  {title} {{On the Transverse-Traceless Projection in Lattice Simulations of
  Gravitational Wave Production}},}\ }\href {\doibase
  10.1088/1475-7516/2011/11/015} {\bibfield  {journal} {\bibinfo  {journal}
  {JCAP}\ }\textbf {\bibinfo {volume} {11}},\ \bibinfo {pages} {015} (\bibinfo
  {year} {2011})},\ \Eprint {http://arxiv.org/abs/1110.0337} {arXiv:1110.0337
  [astro-ph.CO]} \BibitemShut {NoStop}%
\bibitem [{\citenamefont {Buschmann}(2024)}]{buschmann_2024_14285061}%
  \BibitemOpen
  \bibfield  {author} {\bibinfo {author} {\bibfnamefont {Malte}\ \bibnamefont
  {Buschmann}},\ }\href {\doibase 10.5281/zenodo.14285061} {\enquote {\bibinfo
  {title} {Sledgehamr},}\ } (\bibinfo {year} {2024})\BibitemShut {NoStop}%
\bibitem [{\citenamefont {Wainwright}(2012)}]{Wainwright:2011kj}%
  \BibitemOpen
  \bibfield  {author} {\bibinfo {author} {\bibfnamefont {Carroll~L.}\
  \bibnamefont {Wainwright}},\ }\bibfield  {title} {\enquote {\bibinfo {title}
  {{CosmoTransitions: Computing Cosmological Phase Transition Temperatures and
  Bubble Profiles with Multiple Fields}},}\ }\href {\doibase
  10.1016/j.cpc.2012.04.004} {\bibfield  {journal} {\bibinfo  {journal}
  {Comput. Phys. Commun.}\ }\textbf {\bibinfo {volume} {183}},\ \bibinfo
  {pages} {2006--2013} (\bibinfo {year} {2012})},\ \Eprint
  {http://arxiv.org/abs/1109.4189} {arXiv:1109.4189 [hep-ph]} \BibitemShut
  {NoStop}%
\end{thebibliography}%

\clearpage
\appendix
\onecolumngrid

\section{Utility Functions}
\label{sec:utility}
Every nontrivial equation of motion contains some spatial operator such as the gradient operator $\nabla$ or the Laplace operator $\bigtriangleup\equiv \nabla^2$. \sledgehamr\ provides utility functions for the most common versions ready to be used inside kernel functions. In section~\ref{sec:MinimalExample} we have already seen an example for the $\bigtriangleup$ operator:

\begin{lstlisting}[language=c++]
constexpr int order = 2;
double Laplacian_Psi1 = sledgehamr::utils::Laplacian<order>(state, i, j, k, Scalar::Psi1, dx*dx);
\end{lstlisting}
The function arguments should be self-explanatory given the context. The $\bigtriangleup$ operator is calculated through the finite-difference method and \texttt{order} defines the stencil size. Explicitly, \texttt{order} corresponds to the number of ghost cells required for the computation, \textsl{i.e.} \texttt{order=}1, 2, 3 corresponds to the 7-, 13-, 19-point stencil. Note that we want \texttt{order} to be a \texttt{constexpr} as this allows the compiler to always inline the utility function. This will avoid potentially costly function call overhead within the kernel function. For the $\bigtriangleup$ operator, we have implementations for \texttt{order}~$\in\{1,2,3\}$. 

Similarly, field gradients can be computed as
\begin{lstlisting}[language=c++]
constexpr int order = 2;
double grad_x_Psi1 = sledgehamr::utils::Gradient<order>(state, i, j, k, Scalar::Psi1, dx, "x"); 
\end{lstlisting}
Here, the direction spatial direction can be selected through \texttt{"x"}, \texttt{"y"}, and \texttt{"z"} and we have implementations for \texttt{order}~$\in\{1,2,3\}$ corresponding to the three-point, five-point, and seven-point stencil, respectively.

\section{Validation Tests}
As \sledgehamr\ is a relatively mature code insofar as it has already been used in a few publications, tests verifying the convergence of the implemented integrators, operators, and AMR methods have already been performed in the past; see, \textsl{e.g.},~\cite{Buschmann:2021sdq}. Nevertheless, this section will illustrate some basic validation tests for the integrators and spatial operators. We will use the evolution of a plane wave as an example

\begin{equation}
\ddot{u}(\mathbf{x},t) = c^2 \nabla^2 u(\mathbf{x},t),
\end{equation}
due to the existence of a simple analytic solution:
\begin{equation}
    u(\mathbf{x}, t) = A\cos{(\mathbf{k\cdot x}+\omega t + \varphi)}.
\end{equation}
Without a loss of generality, we choose $A=c=1$, $\varphi=0$, $\omega=\pi$, and 
\begin{equation}
\mathbf{k} = \pi\begin{pmatrix}\cos\theta\cos\phi\\\sin\phi\\-\cos\theta\sin\phi\end{pmatrix}.
\end{equation}

With this choice of wavevector $\mathbf{k}$, the system corresponds to a plane wave moving in $x'$-direction, where $x'$ is the $x$-direction rotated slightly around the $y$- and $z$-axis by angle $\theta$ and $\phi$, respectively. We do this rotation to misalign the travel direction of the wave with respect to the underlying 3D grid. This way we can probe all three components of the spatial operator simultaneously. However, we must choose the rotation angles carefully as we are restricted to a simulation box with equal side lengths and periodic boundary conditions. To fulfill periodic boundary conditions, the box length needs to be a multiple of the wavelength projected onto the x-, y-, and z-axis, $\lambda_i=2\pi/k_i$, and thus, the rotation angles cannot be too small, or the resulting simulation volume would need to be enormous.

We opted for the choice $\theta=\arctan 1/2\sim 26.6^{\circ}$ and $\phi=\arcsin{3/5}\sim 36.7^{\circ}$. In this case, we find a reasonably small simulation volume with $L=\text{LCM}_i(\lambda_i)=10\sqrt{5}$ that fits exactly eight, five, and six wave packets along each axis. We simulate from $t_0=0$ to $t_1=100 t_p$, where $t_p=2\pi/\omega$ is the Poincaré recurrence time. We run a variety of simulations using a fourth-order spatial Laplacian and $\Delta t=0.32 \Delta x$. The simulations differ in the grid size from $32^3$ to $512^3$ cells and use different integrators, specifically the third- and fourth-order Runge-Kutta (RK) integrator and fourth- and fifth-order Runge-Kutta-Nyström (RKN) integrator.

\begin{figure*}[!htb]
\centering
\includegraphics[width=0.5\textwidth]{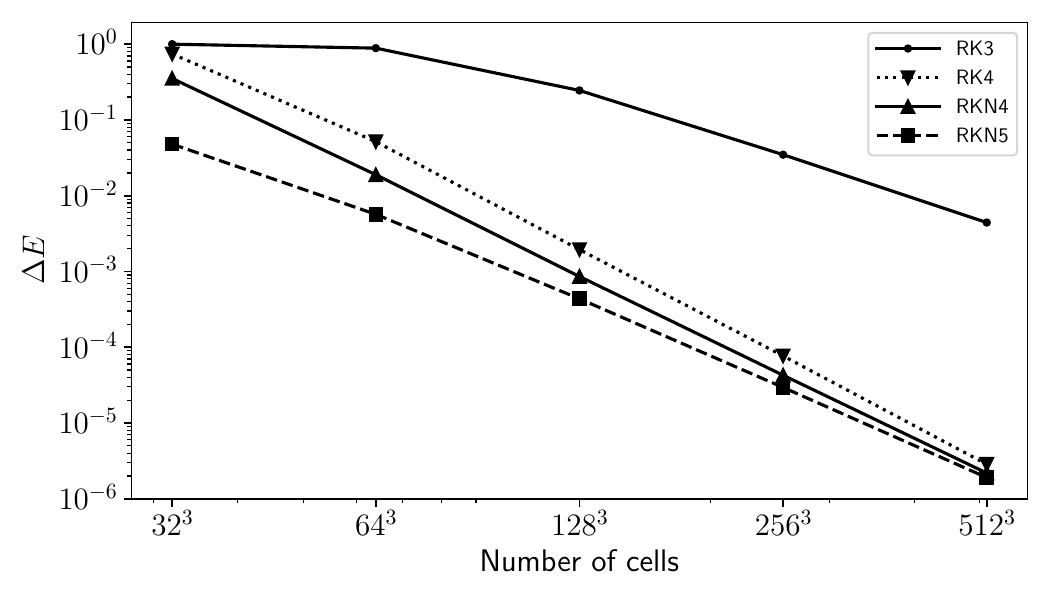}
\caption{Difference in energy $\Delta E$ between simulation and analytic solution of a plane wave for a few different integrator schemes and lattice sizes.}
\label{fig:validation}
\end{figure*}

For each of these simulations, we measure the level of convergence by comparing the total wave energy $E=\dot{u}^2/2 + (\nabla u)^2/2$ at the final state with the analytic expectation of $E_{\rm{exact}}=\pi^2/2$. We present the difference $\Delta E= E - E_{\rm{exact}}$ in Fig.~\ref{fig:validation}. Obviously, the convergence improves for larger grids and higher-order integrators. We verified that the rate of improvement matches expectations by fitting a power-law function to each curve of a given integrator. Here, we excluded all data points where $\Delta E\sim \mathcal{O}(1)$ since the level of convergence is too poor in these runs and thus not meaningful. We find a power-law index of $-3.0$ for RK3, $-4.7$ for RK4, $-4.3$ for RKN4, and $-4.0$ for RKN5, which matches the expectation that the index should be roughly $-p$ or better for an integrator of order $p$. This relation does not hold for the fifth-order RKN5 algorithm as these simulations are limited by the fourth-order Laplacian, explaining the shallower slope of $-4.0$. We thus conclude that the level of convergence matches expectations.

\section{Kreiss-Oliger dissipation}
\label{sec:AppKO}
\begin{figure*}[!htb]
\centering
\includegraphics[width=0.59\textwidth]{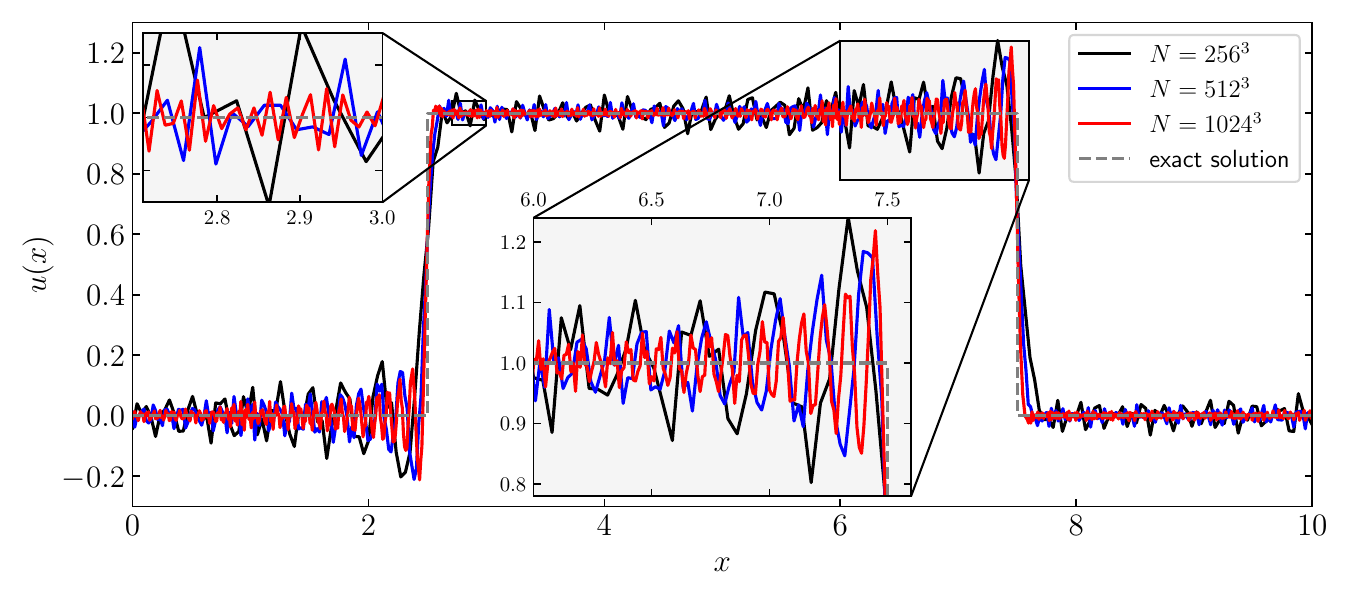}
\includegraphics[width=0.4\textwidth]{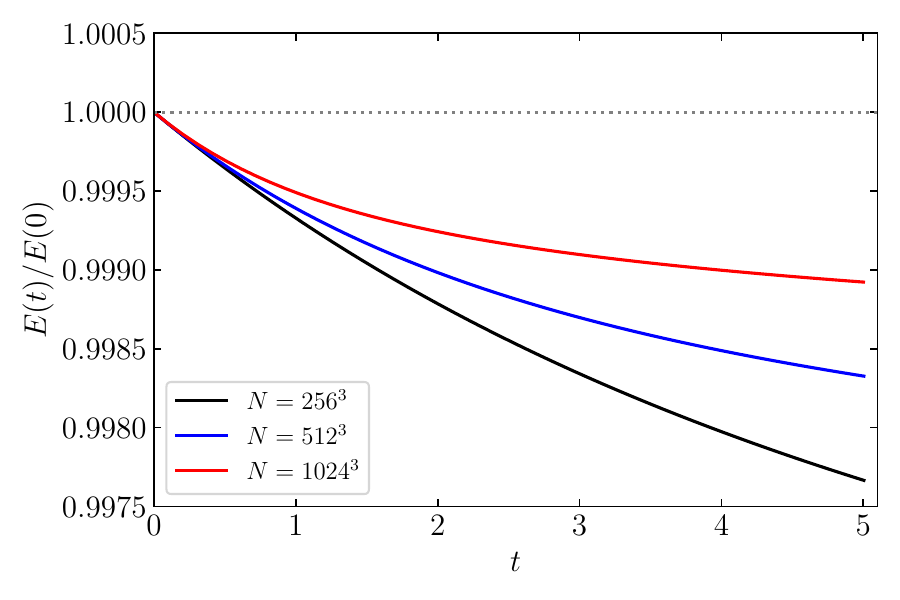}\\
\includegraphics[width=0.59\textwidth]{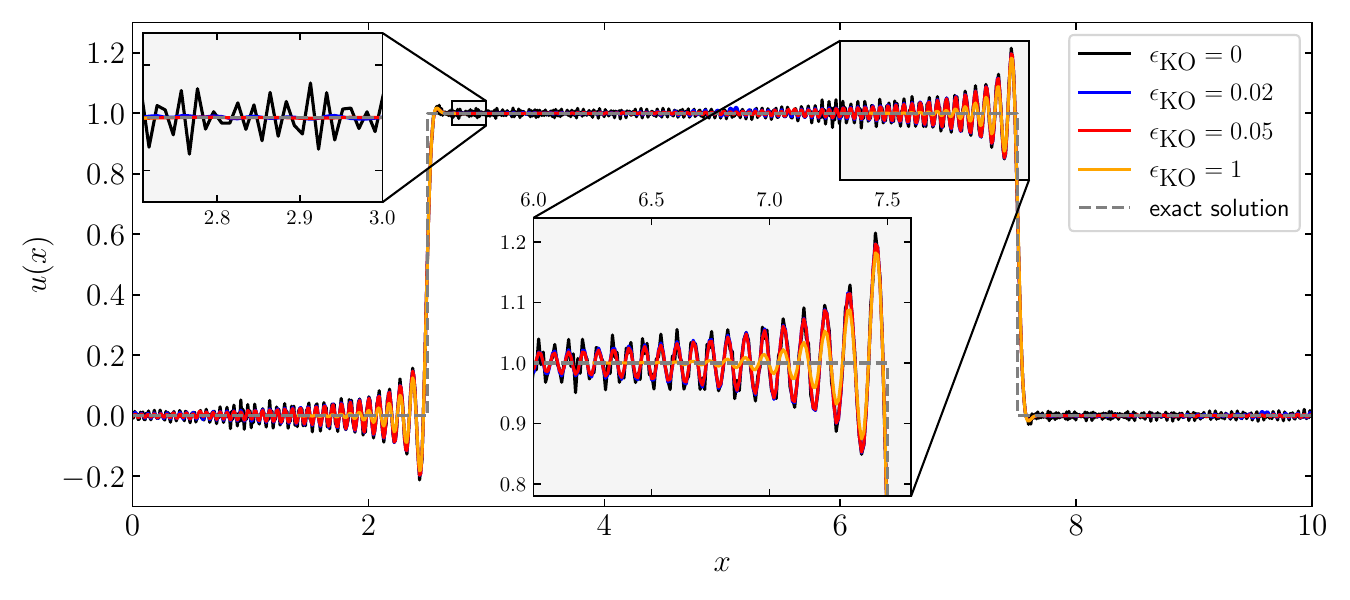}
\includegraphics[width=0.4\textwidth]{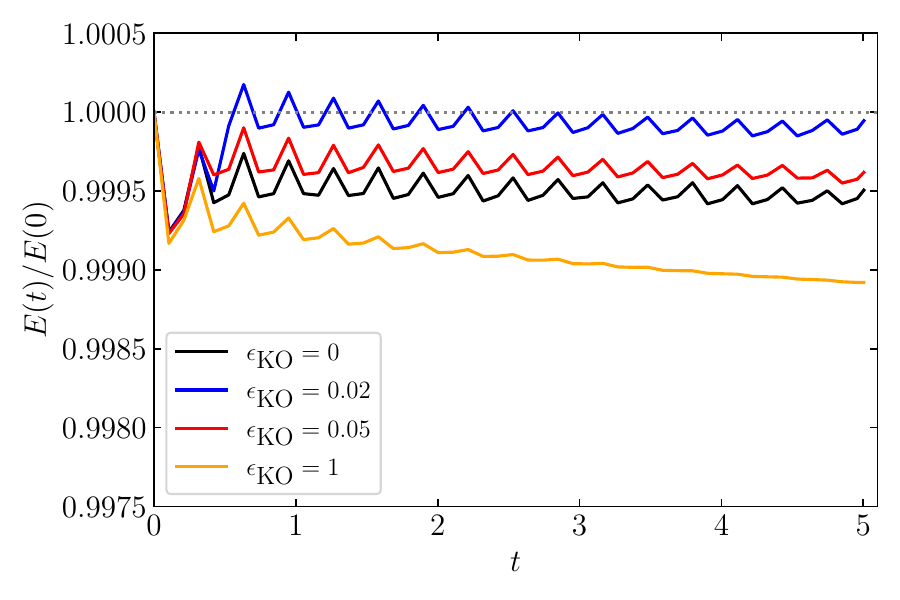}
\caption{Results of simulations of a propagating discontinuous wave. The left column shows a slice through the field $u$ along the x-axis at the end of the simulation, whereas the right column illustrates the level of energy conservation as a function of time. The top row is a set of simulations using neither AMR nor Kreiss-Oliger dissipation but utilizing various grid resolutions. The bottom row contains simulations using AMR in conjunction with Kreiss-Oliger dissipation.}
\label{fig:KO}
\end{figure*}

The purpose of Kreiss-Oliger dissipation is to damp high-frequency numerical noise, primarily originating at the intersection between coarse and fine-level grids. To illustrate the effects of this extra dissipation term, we perform a dedicated study of a propagating discontinuous wave described by the steady-flow advection equation
\begin{equation}
    \frac{\partial u}{\partial t} + \mathbf{c}\cdot \nabla u = 0
\end{equation}
with wave speed $\bf{c}=(2,2,2)^T$. If such a system is solved with an Runge-Kutta algorithm and finite-difference techniques, the solution will develop unphysical high-frequency oscillations at the discontinuity similar to the artificial noise appearing at a coarse-fine boundary. We will use a fourth-order Runge-Kutta integrator and an initial state with a box profile that fills out half of the simulation volume. With a box size of $L=10$, we simulate from $t=0$ to $t=5$ with $\Delta t=0.3 \Delta x$, which means the wave propagates through the entire simulation box exactly once. We compute gradients with a five-point stencil and the Kreiss-Oliger dissipation term with a seven-point stencil. We present the results of simulations with different base resolutions, with and without AMR, and different values of $\epsilon_\text{KO}$ in Fig.~\ref{fig:KO}.

We start with the simplest case of a static grid without AMR and no Kreis-Oliger dissipation. These are shown in the top row of Fig.~\ref{fig:KO} and clearly highlight the emergence of high-frequency noise near the discontinuity. The frequency of this noise depends on the grid spacing since it originates from the finite-difference gradients. While the amplitude of the noise decreases with increasing resolution, the actual improvement is unsurprisingly relatively minor. Noise is still evident even with 1,024$^3$ grid cells. Energy conservation is at the $\mathcal{O}(0.1-0.3)\%$ level.

The bottom row of Fig.~\ref{fig:KO} contains simulations with a $256^3$ coarse-level resolution and up to two refinement levels based on a truncation error threshold of $\epsilon_u=10^{-3}$. We also utilize a variety of different levels of Kreiss-Oliger dissipation in these simulations, which allows us to make the following observations: We first note that the noise frequency corresponds to that of the finest refinement level with $N=1024$ cells since the area around the discontinuity is refined to the maximum extend. Without Kreiss-Oliger dissipation, $\epsilon_\text{KO}=0$, the properties of the noise are overall similar to that of the static 1,024$^3$ grid cells' simulation. Note, however, that the energy conservation of the AMR simulation has improved over the static simulation. This is a side effect of mixing grids with different resolutions and thus computing finite-difference gradients using a stencil that spans a different length on each level. This causes an averaging effect at the level boundaries that limits the impact of the numerical noise on the energy loss.

Crucially though and despite better energy conservation, numerical noise is still very visible throughout the entire simulation volume if no Kreiss-Oliger dissipation is used. With Kreiss-Oliger dissipation, however, even small values of $\epsilon_\text{KO}$ of $\mathcal{O}(10^{-2})$ are sufficient to completely eradicate any noise on the far side of the discontinuity. As a side effect, the simulation will also run faster as the noiseless regime will often be refined at a lesser level. Substantial values of $\epsilon_\text{KO}$ of $\mathcal{O}(1)$ will dampen the noise even closer to the discontinuity; however, this will also cause visible leakage in energy. In practice, we find that $\epsilon_\text{KO}$ of $\mathcal{O}(10^{-2})$ is often sufficient to suppress the noise emerging at coarse-fine boundaries without significantly harming energy conservation.

\section{Picking the right truncation error threshold}
One of \sledgehamr's key parameters is the truncation error threshold $\epsilon_\phi$ as it is one of the main controls for where and how much of the lattice is refined. Section~\ref{sec:tagging} describes how truncation error estimates are computed and how the threshold $\epsilon_\phi$ is used to tag cells for refinement. In this section, we will go over how $\epsilon_\phi$ is related to the numerical convergence of the simulation and how it affects the amount of computational resources required. While the quantitative results will depend on the exact physics scenario, this section will nevertheless be a good illustration of the general correlation between the threshold and the quality of the solution.

As in section~\ref{sec:tagging}, we will use a scalar field $\phi(x)$ undergoing a phase transition from a false vacuum to a true vacuum as an example. The following results can be reproduced with the \texttt{FirstOrderPhaseTransition} project included in \sledgehamr. Explicitly, we solve the equation of motion
\begin{align}
    \frac{\partial^2\phi}{\partial t^2} = \nabla^2 \phi - \frac{\partial V}{\partial \phi}
\end{align}
with the double-well potential
\begin{align}
V(\phi) =  \frac{1}{2} \phi^2 - \frac{\left(\sqrt{9-8 \bar \lambda }+3\right)}{4 \bar \lambda } \phi^3 + \frac{\left(\sqrt{9-8  \bar \lambda }+3\right)^2}{32 \bar \lambda } \phi^4,
\end{align}
where we chose $\bar\lambda=0.84$. With this choice of potential, the field $\phi(x)$ will transition from a false vacuum at $\phi(x)=0$ to the true vacuum at $\phi(x)=1$. The transition requires a nucleation point to get started, for which we will use the code package \texttt{CosmoTransitions}~\cite{Wainwright:2011kj} to generate a single isolated nucleation bubble as an initial state. This setup is identical to what is shown in Fig.~\ref{fig:truncation_error} with the only difference being that it features two of those bubbles rather than just one.

We will use a fourth-order Runge-Kutta-Nyström integrator with $\Delta t=0.3\Delta x$ and a coarse-level lattice size of $256^3$ cells. The Laplacian is solved using the 13-point finite-difference stencil. The simulation volume is a box with side length $L=200$, which is significantly larger than the initial bubble radius of just $r(t=0)=6.25$. We simulate until $t=75$, at which point the bubble will have filled out a significant portion of the simulation volume. By default, truncation errors are absolute quantities, and to account for a change in scale, we normalize them using the framework outlined in section~\ref{sec:ModTE}. Explicitly, we use the transformation $\Delta \phi_\text{cf}\rightarrow \Delta\phi_\text{cf}/\text{max}(\phi)$ and $\Delta (\partial\phi/\partial t)_\text{cf}\rightarrow \Delta(\partial\phi/\partial t)_\text{cf}/\text{max}(\partial\phi/\partial t)$. We repeat the simulation for a few different values of $\epsilon_\phi\equiv\epsilon_{\partial\phi/\partial t}$. For consistency, all simulations are run on four GPU nodes with a total of 16 NVIDIA A100 GPUs.

\begin{figure*}[!htb]
\centering
\includegraphics[width=1\textwidth]{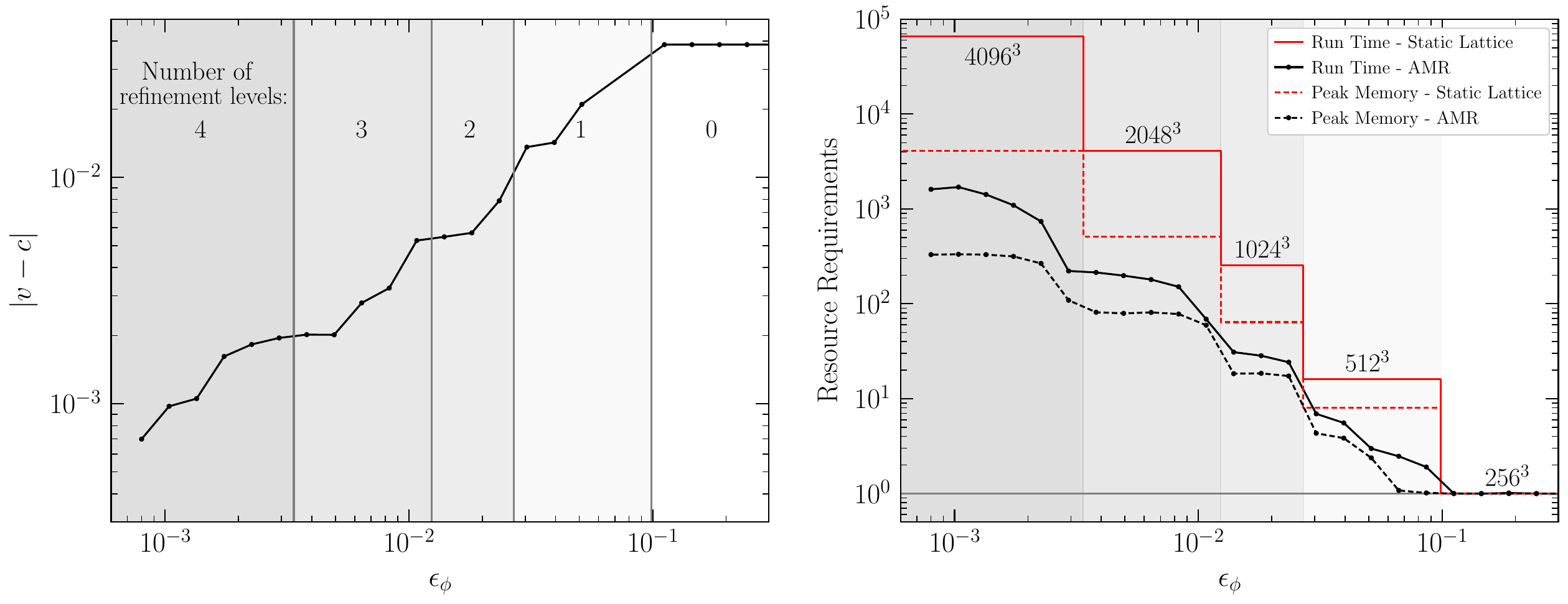}
\caption{\textsl{(left)} Absolute difference between the bubble wall velocity $v$ and its expected value $c=1$. As the truncation error threshold $\epsilon_\phi$ is lowered, an increasing number of refinement levels are added, thus improving the quality of the solution and reducing the discrepancy between $v$ and $c$. \textsl{(right)} Resource requirements for AMR simulations with different $\epsilon_\phi$, normalized to a static $256^3$ lattice sim. The results are compared in red to the expected resource consumption of a static lattice matching the finest resolution of the respective AMR simulation.}
\label{fig:truncation_scaling}
\end{figure*}

The bubble should expand at the speed of light, which is $c=1$ with the above field definition. However, the bubble wall is notoriously sensitive to discretization effects, and insufficient resolution will slow down the bubble expansion. As a metric for the quality of the solution, we, therefore, extract the location of the bubble wall at $\phi(x)=0.5$ at every coarse-level time step and use this to measure the wall velocity $v$ through a linear fit at $t>50$. The result for the difference $|v-c|$ is shown in Fig.~\ref{fig:truncation_scaling}. As expected, reducing the threshold reduces the discrepancy: The bubble velocity is about 4\% slower than it should be for $\epsilon_\phi>0.1$. With such a high threshold, no refinement is triggered at any point during the simulation. For threshold values below $0.1$, an increasing amount of refinement is introduced, and, subsequently, the quality of the solution improves continuously. By $\epsilon_\phi\sim10^{-3}$ the bubble velocity differs only by $\sim0.1\%$ from the expected value. This depends strongly on the actual physics scenario, but we often find threshold values between $10^{-4}$ and a few$\times10^{-3}$ to work well, although your mileage may vary.

Of course, introducing a large amount of refinement increases the computational cost both in terms of computation time and memory consumption. Since we are often resource limited, this can typically be the better metric when setting up a simulation in practice. To illustrate the effect of $\epsilon_\phi$ on those quantities, we keep track of the total run-time and the peak memory consumption in our simulations. For a fair comparison, the run-time excludes I/O operations and the initialization period but includes the entire regridding procedure. We show the results in Fig.~\ref{fig:truncation_scaling} where all quantities are normalized to the unrefined $256^3$ grid. We compare the results with the expected resource consumption of a static non-AMR lattice simulation assuming that it matches the resolution of the finest refinement level of the AMR simulation at a given $\epsilon_\phi$. Note that we did not run any large static simulations, but used the unrefined $256^3$ simulation as a benchmark. To extrapolate to larger lattice sizes, we assume that the peak memory consumption scales like $N_g^3$ and run-time scales like $N_g^3 N_t$, where $N_g^3$ is the total number of lattice sites, and $N_t$ the number of time steps. Of course, this scaling is very naïve in practice but will suffice for this simple comparison.

Unsurprisingly, an AMR simulation is more efficient than a static lattice in both run-time and memory consumption. 
Run-time increases by roughly a factor of $\sim6$ for every additional refinement level, and peak memory consumption by a factor of $\sim 4$. This can be compared to the respective factors of $\sim 16$ and $\sim 8$ if a static lattice resolution is increased by the same amount. At $\epsilon_\phi\sim10^{-3}$, peak memory consumption is lower by a factor of $12$ and run-time is lower by a factor of $40$ compared to a static simulation. Note that we did not make any attempts at optimizing the setup of this particular set of simulations. This scaling is highly problem-specific as well: Bubble walls form 2D surfaces in 3D spaces, whereas, $e.g.$ cosmic strings in Fig.~\ref{fig:axion_string} are just 1D objects. This massively influences how much volume is refined per refinement level and thus how much additional resources are required.

Lastly, we want to point out that memory consumption and run time can be balanced against each other, which is most easily done by changing the blocking factor. In this particular suite of simulations, we used a blocking factor of 32, which is relatively large for a coarse level of $256^3$ cells. If we were to reduce the blocking factor, the refined volume would be generally smaller as the refined grid can fit more tightly around tagged cells. This reduces the peak memory consumption. However, that comes at the price of introducing an increasing amount of coarse-fine and internal boundaries, which are computationally expensive to deal with. It increases the number of ghost cells that need to be interpolated at every intermediate time step, and the net result is often an increase in computation time despite featuring less lattice sites overall.

%Frequency of noise
%Energy conservation is better since AMR
%Noise still exists at the 1,024^3 level buTo illustrate tht minor KO reduces it already massively. Improves energy conservation. 
%Large KO improves solution visibly in this extreme scenario but also dissipates energy quite noticeably. Small KO of 0.01 level seems to be working well.
% Reduced AMR coverage.}
\section{List of run-time parameters}
\label{sec:parameters}
Below is a list of all available run-time parameters including their default values, loosely separated into five different categories: \texttt{sim.*}, \texttt{amr.*}, \texttt{integrator.*}, \texttt{input.*}, and \texttt{output.*}. Additionally, we have the parameter \texttt{project.name} to select the project to run. Default values are used if the parameter has not been specified in the \texttt{inputs} file. If no default value is given, the parameter in question is mandatory and has to be added to the \texttt{inputs} file.

\begin{table}[ht!]
\centering
\renewcommand{\arraystretch}{1.4}
\begin{tabularx}{1.0\textwidth}{|l l l X|}
\hline
\texttt{sim}.*     & Range          & Default  & Description \\\hline
\texttt{t\_start}  & $\mathbb{R}$   &           & Start time $t_\text{start}$.\\
\texttt{t\_end}    & $\mathbb{R}$   &           & Final time $t_\text{end}$.\\
\texttt{L}         & $\mathbb{R}^+$ &           & Box length $L$.\\
\texttt{cfl}       & $\mathbb{R}^+$ &           & Sets $\Delta t_\ell/\Delta x_\ell$ for the Courant-Friedrichs-Lewy condition. This is used to derive the time step size $\Delta t_\ell$ since $\Delta x_\ell$ is uniquely determined through \texttt{sim.L} and \texttt{amr.coarse\_level\_grid\_size}. See section~\ref{sec:AMR}.\\
\texttt{gravitational\_waves} & $\{0,1\}$ & $0$   & Whether to run with gravitational waves. See section~\ref{sec:GW}.\\
\texttt{dissipation\_strength} & $\mathbb{R}^+$ & 0  & Sets the strength $\epsilon_\text{KO}$ of Kreiss-Oliger dissipation for all field components. This value will be used as a default value for all field components. If set to $0$ the dissipation term will be removed. See section~\ref{sec:KO}.\\
\texttt{dissipation\_strength\_\$ScalarField} & $\mathbb{R}^+$ & $\epsilon_\text{KO}$  & Sets the dissipation strength $\epsilon_\text{KO}$ for a particular field component by overwriting the value set by \texttt{sim.dissipation\_strength}.\\
\texttt{dissipation\_order} & $\{2,3\}$ & \texttt{Null} & Defines the order $p$ of Kreiss-Oliger dissipation. The actual value corresponds to the number of ghost cells needed for the computation. This parameter needs to be set only if $\epsilon_\text{KO}>0$ for any field component. See section~\ref{sec:KO}.\\
\hline
\end{tabularx}
\caption{General simulation parameters.}
\label{tab:sim}
\end{table}

\begin{table}[ht!]
\centering
\renewcommand{\arraystretch}{1.4}
\begin{tabularx}{1.0\textwidth}{|l l l X|}
\hline
\texttt{input.*}     & Range          & Default  & Description \\\hline
\texttt{initial\_state}  & string & "" & Path to hdf5 file containing the initial state. It is also possible to provide the path to a checkpoint, in which case the level data will be initialized from the checkpoint. However, no other metadata provided by the checkpoint will be used to initialize the simulation. See section~\ref{sec:checkpoints}. \\
\texttt{initial\_state\_\$ScalarField}  & string & "" & Path to hdf5 file containing the initial state for the scalar field \texttt{\$ScalarField}. See section~\ref{sec:input}. \\
\texttt{upsample}  & $2^\mathbb{N}$ & 1 & If \texttt{upsample} $>1$ the provided initial state will be up-scaled by the given factor. It is assumed that the provided initial state is of size \texttt{coarse\_level\_grid\_size}/\texttt{upsample}.\\
\texttt{restart}  & $\{0,1\}$ & 0 & If $1$ we will restart the simulation from a checkpoint. By default, the latest checkpoint within the output folder will be selected. \\
\texttt{select\_checkpoint}  & string or $\mathbb{N}$ & \texttt{Null} & With this parameter, we can manually select a checkpoint from which to restart the simulation. If \texttt{select\_checkpoint} $\in\mathbb{N}$ the corresponding checkpoint in the output folder will be used. Alternatively, \texttt{select\_checkpoint} can be set to the path of a checkpoint.\\
\texttt{delete\_restart\_checkpoint} & $\{0,1\}$ & 0 & Defines whether we keep or delete the checkpoint from which we restarted the simulation.\\
\texttt{get\_box\_layout\_nodes}  & $2^\mathbb{N}$ & 0 & If \texttt{get\_box\_layout\_nodes} $>0$ no simulation will be performed. Instead, the coarse-level box layout will be written to an hdf5 file corresponding to the number of nodes set by this parameter. This can be used to set up an initial state split into chunks. See section~\ref{sec:input}.\\
\hline
\end{tabularx}
\caption{Initial state parameters.}
\label{tab:input}
\end{table}

\begin{table}[ht!]
\centering
\renewcommand{\arraystretch}{1.4}
\begin{tabularx}{1.0\textwidth}{|l l l X|}
\hline
\texttt{amr.*}     & Range          & Default  & Description \\\hline
\texttt{coarse\_level\_grid\_size}  & $2^\mathbb{N}$ &      & Number of cells at the coarse level in each spatial direction.\\
\texttt{max\_refinement\_levels}  & $\mathbb{N}$ &           & Total number of refinement levels. If set to $0$ AMR will effectively be switched off.\\
\texttt{blocking\_factor}  & $2^\mathbb{N}$ & 8          & The length of each box is a multiple of the blocking factor.\\
\texttt{nghost}  & $\mathbb{N}$ & 0        & Number of ghost cells. Must be smaller than the blocking factor.\\
\texttt{n\_error\_buf}  & $\mathbb{N}$ & 1          & Number of buffer cells during a regrid. Must be smaller than the blocking factor.\\
\texttt{regrid\_dt}  & $\mathbb{R}^+$ & \texttt{DBL\_MAX} & Regrid interval $\Delta t_\text{regrid}^{\ell=0}$ at the coarse level, where $\Delta t_\text{regrid}^{\ell}$ = $\Delta t_\text{regrid}^{\ell=0}/2^\ell$.\\
\texttt{force\_global\_regrid\_at\_restart}  & $\{0,1\}$ & 0 & If set to $1$ a global regrid will be performed when (re-)starting a simulation instead of a local regrid. If 0, a local regrid will be attempted first. \\
\texttt{tagging\_on\_gpu}  & $\{0,1\}$ & 0 & The cell tagging procedure during a regrid is usually done on CPUs even if GPUs are available. This is mostly to keep track of tagging statistics for diagnostic purposes. The tagging procedure is rarely a bottleneck, so not much time is wasted by running it on CPUs. However, with this toggle, we can force it to run in GPUs in lieu of collecting statistics. \\
\texttt{max\_local\_regrids}  & $\mathbb{N}$ & 10 & Caps the number of consecutive local regrids. If \texttt{max\_local\_regrids} are reached, a global regrid will be performed instead, and the counter resets.\\
\texttt{volume\_threshold\_strong}  & $\mathbb{R}^+$ & 1.1 & $\epsilon_s$ threshold during a local regrid. See section~\ref{sec:regrid}.\\
\texttt{volume\_threshold\_weak}  & $\mathbb{R}^+$ & 1.05 & $\epsilon_w$ threshold during a local regrid. See section~\ref{sec:regrid}.\\
\texttt{interpolation\_type}  & $\{0,1,2,4\}$ & 4          & Selects what type of spatial interpolation should be used to fill ghost cells.\newline 0 = Piecewise constant interpolation\newline 1 = Linear conservative interpolation\newline 2 = Quadratic polynomial interpolation\newline 4 = Quartic polynomial conservative interpolation\\
\texttt{te\_crit}  & $\mathbb{R}^+$ & \texttt{DBL\_MAX} & Sets the TEE threshold $\epsilon_\phi$ for all scalar fields. See section~\ref{sec:tagging}.\\
\texttt{te\_crit\_\$ScalarField}  & $\mathbb{R}^+$ & $\epsilon_\phi$ & Sets the TEE threshold $\epsilon_\phi$ for a particular field component by overwriting the value set by \texttt{te\_crit}.\\
\texttt{semistatic\_sim}  & $\{0,1\}$ & 0 & Special mode where no AMR is enabled. However, each time the virtual function \texttt{CreateLevelIf} returns \texttt{true}, the entire simulation volume is refined by a factor of 2. Only the refined data is kept from there onwards.\\
\texttt{increase\_coarse\_level\_resolution}  & $\{0,1\}$ & 0 & Occasionally, it is beneficial to change the coarse-level resolution. When this parameter is set to 1 the coarse-level resolution is increased by a factor of 2 during the next run. For any subsequent runs, we will need to set \texttt{increase\_coarse\_level\_resolution} back to 0 to not increase the coarse-level resolution a second time and readjust the input parameters such as \texttt{coarse\_level\_grid\_size} and \texttt{regrid\_dt} accordingly.\\
\hline
\end{tabularx}
\caption{AMR parameters.}
\label{tab:amr}
\end{table}

\begin{table}[ht!]
\centering
\renewcommand{\arraystretch}{1.4}
\begin{tabularx}{1.0\textwidth}{|l l l X|}
\hline
\texttt{output.*}     & Range          & Default  & Description \\\hline
\texttt{output\_folder}  & string &  & All simulation output will be written to this path. If the folder does not exist yet, it will be created.\\
\texttt{alternative\_output\_folder}  & string & \texttt{Null} & A secondary output folder. It is possible to alternate output between primary and secondary output folders; see \texttt{output.\$OutputType.alternate}. This can help alleviate disk quota constraints. \\
\texttt{rename\_old\_output}  & $\{0,1\}$ & 0 & If the output folder already exists but we are not explicitly restarting a sim with \texttt{input.restart} $=1$, then the simulation run will be aborted. This is to avoid accidentally overwriting existing simulation output. If \texttt{rename\_old\_output} $=1$, however, the existing output folder will be renamed by adding a unique string. The simulation will then continue as normal.\\
\texttt{output\_of\_initial\_state}  & $\{0,1\}$ & 1 & If $1$, we will immediately start producing output at $t=t_\text{start}$. If $0$, we will wait for one write interval.\\
\texttt{write\_at\_start}  & $\{0,1\}$ & 0           & If $1$, we will immediately start producing output when restarting from a checkpoint. If $1$, we will wait for one write interval since the time at which the respective output has been written last.\\
\texttt{\$OutputType.interval}  & $\mathbb{R}$ & -1 & Sets the write interval. If negative, the respective output type will be disabled. \\
\texttt{\$OutputType.alternate}  & $\{0,1\}$ & 0 & If 1, the output will alternate between the primary and secondary output folders. \\
\texttt{\$OutputType.min\_t}  & $\mathbb{R}$ & \texttt{-DBL\_MAX} & No output will be written if $t < t_\text{min}$.\\
\texttt{\$OutputType.max\_t}  & $\mathbb{R}$ & \texttt{ DBL\_MAX} & No output will be written if $t > t_\text{max}$.\\
\texttt{checkpoints.rolling}  & $\{0,1\}$ & 0 & If $1$, only the last checkpoint is being kept on disk. Any preceding checkpoint will be deleted.\\
\texttt{gw\_spectra.projection\_type}  & $\{1,2\}$ & 2 & Sets the order of the projector in Eq.~\ref{eq:Lambda_ijlm}, see section~\ref{sec:gwspectra}.\\
\texttt{projections.max\_level}  & $\mathbb{N}$ & \texttt{INT\_MAX} & Sets the maximum level to be included in the projection output; see section~\ref{sec:projections}.\\
\texttt{\$BoxOutput.downsample\_factor}  & $2^\mathbb{N}$ & 1 & All output types that save a 3D volume (\texttt{\$BoxOutput=\{coarse\_box}, \texttt{coarse\_box\_truncation\_error}, \texttt{full\_box}, \texttt{full\_box\_truncation\_error\}}) can be downsampled by this factor.\\
\hline
\end{tabularx}
\caption{Simulation output. Here, \texttt{\$OutputType} is any of the following: \texttt{checkpoints},
\texttt{slices},
\texttt{coarse\_box},
\texttt{full\_box},
\texttt{slices\_truncation\_error},
\texttt{coarse\_box\_truncation\_error},
\texttt{full\_box\_truncation\_error},
\texttt{projections},
\texttt{spectra},
\texttt{gw\_spectra},
\texttt{performance\_monitor}, or
\texttt{amrex\_plotfile}.}
\label{tab:output}
\end{table}

\begin{table}[ht!]
\centering
\renewcommand{\arraystretch}{1.4}
\begin{tabularx}{1.0\textwidth}{|l l l X|}
\hline
\texttt{integrator.*}     & Range          & Default  & Description \\\hline
\texttt{type}  & $\{0-4,10,20-22\}$ & & Selects the integrator type.\newline  
0 = \amrex\ integrator - User defined Runge-Kutta (RK) Butcher Tableau\newline 
1 = \amrex\ integrator - Forward Euler\newline
2 = \amrex\ integrator - Trapezoid Method\newline
3 = \amrex\ integrator - SSPRK3 Method\newline
4 = \amrex\ integrator - RK4 Method\newline
10 = Low-storage SSPRK3 Method (smaller memory footprint than SSPRK3)\newline
20 = User-defined Runge-Kutta-Nyström (RKN) Butcher Tableau\newline
21 = RKN4 Method\newline
22 = RKN5 Method\\
\texttt{rk.weights} & $n\times\mathbb{R}$ & \texttt{Null} & RK Butcher tableau weights provided as a list of values. Needed if \texttt{integrator.type=0}.\\
\texttt{rk.nodes} & $n\times\mathbb{R}$ & \texttt{Null} & RK Butcher tableau nodes provided as a list of values. Needed if \texttt{integrator.type=0}.\\
\texttt{rk.tableau} & $n\times\mathbb{R}$& \texttt{Null} & RK Butcher tableau. Shall be the flattened lower triangular matrix including the diagonal. Needed if \texttt{integrator.type=0}. \\
\texttt{rkn.weights\_b} & $n\times\mathbb{R}$ & \texttt{Null} & RKN Butcher tableau weights $\mathbf{b}$ provided as a list of values. Needed if \texttt{integrator.type=20}.\\
\texttt{rkn.weights\_bar\_b} & $n\times\mathbb{R}$ & \texttt{Null} & RKN Butcher tableau weights $\mathbf{\bar{b}}$ provided as a list of values. Needed if \texttt{integrator.type=20}. \\
\texttt{rkn.nodes} & $n\times\mathbb{R}$ & \texttt{Null} & RKN Butcher tableau nodes provided as a list of values. Needed if if \texttt{integrator.type=20}.\\
\texttt{rkn.tableau} & $n\times\mathbb{R}$& \texttt{Null} & RKN Butcher tableau. Shall be the flattened lower triangular matrix including the diagonal. Needed if \texttt{integrator.type=20}.\\
\hline
\end{tabularx}
\caption{Integrator settings.}
\label{tab:intergrator}
\end{table}

\end{document}